\input harvmac

\input amssym
\input epsf


\newfam\frakfam
\font\teneufm=eufm10
\font\seveneufm=eufm7
\font\fiveeufm=eufm5
\textfont\frakfam=\teneufm
\scriptfont\frakfam=\seveneufm
\scriptscriptfont\frakfam=\fiveeufm


\def\bb{
\font\tenmsb=msbm10
\font\sevenmsb=msbm7
\font\fivemsb=msbm5
\textfont1=\tenmsb
\scriptfont1=\sevenmsb
\scriptscriptfont1=\fivemsb
}


\newfam\dsromfam
\font\tendsrom=dsrom10
\textfont\dsromfam=\tendsrom
\def\ds{\fam\dsromfam \tendsrom}


\newfam\mbffam
\font\tenmbf=cmmib10
\font\sevenmbf=cmmib7
\font\fivembf=cmmib5
\textfont\mbffam=\tenmbf
\scriptfont\mbffam=\sevenmbf
\scriptscriptfont\mbffam=\fivembf


\newfam\mbfcalfam
\font\tenmbfcal=cmbsy10
\font\sevenmbfcal=cmbsy7
\font\fivembfcal=cmbsy5
\textfont\mbfcalfam=\tenmbfcal
\scriptfont\mbfcalfam=\sevenmbfcal
\scriptscriptfont\mbfcalfam=\fivembfcal


\newfam\mscrfam
\font\tenmscr=rsfs10
\font\sevenmscr=rsfs7
\font\fivemscr=rsfs5
\textfont\mscrfam=\tenmscr
\scriptfont\mscrfam=\sevenmscr
\scriptscriptfont\mscrfam=\fivemscr



\def\vev#1{\left\langle #1\right\rangle}


\def\tilde{\widetilde}

\def\t{\tilde}
\def\hat{\widehat}
\def\h{\hat}
\def\bar{\overline}
\def\b{\bar}
\def\bsq#1{{{\b{#1}}^{\lower 2.5pt\hbox{$\scriptstyle 2$}}}}
\def\bexp#1#2{{{\b{#1}}^{\lower 2.5pt\hbox{$\scriptstyle #2$}}}}
\def\dotexp#1#2{{{#1}^{\lower 2.5pt\hbox{$\scriptstyle #2$}}}}

\def\IL{\relax{\rm I\kern-.18em L}}
\def\IH{\relax{\rm I\kern-.18em H}}
\def\IR{\relax{\rm I\kern-.18em R}}
\def\IC{\relax{\rm I\kern-0.54 em C}}

\def\rt2{\sqrt{2}}
\def\half {{1 \over 2}}


\font\tenbifull=cmmib10
\font\tenbimed=cmmib7
\font\tenbismall=cmmib5
\textfont9=\tenbifull \scriptfont9=\tenbimed
\scriptscriptfont9=\tenbismall

\mathchardef\bbGamma="7000
\mathchardef\bbDelta="7001
\mathchardef\bbPhi="7002
\mathchardef\bbAlpha="7003
\mathchardef\bbXi="7004
\mathchardef\bbPi="7005
\mathchardef\bbSigma="7006
\mathchardef\bbUpsilon="7007
\mathchardef\bbTheta="7008
\mathchardef\bbPsi="7009
\mathchardef\bbOmega="700A
\mathchardef\bbalpha="710B
\mathchardef\bbbeta="710C
\mathchardef\bbgamma="710D
\mathchardef\bbdelta="710E
\mathchardef\bbepsilon="710F
\mathchardef\bbzeta="7110
\mathchardef\bbeta="7111
\mathchardef\bbtheta="7112
\mathchardef\bbiota="7113
\mathchardef\bbkappa="7114
\mathchardef\bblambda="7115
\mathchardef\bbmu="7116
\mathchardef\bbnu="7117
\mathchardef\bbxi="7118
\mathchardef\bbpi="7119
\mathchardef\bbrho="711A
\mathchardef\bbsigma="711B
\mathchardef\bbtau="711C
\mathchardef\bbupsilon="711D
\mathchardef\bbphi="711E
\mathchardef\bbchi="711F
\mathchardef\bbpsi="7120
\mathchardef\bbomega="7121
\mathchardef\bbvarepsilon="7122
\mathchardef\bbvartheta="7123
\mathchardef\bbvarpi="7124
\mathchardef\bbvarrho="7125
\mathchardef\bbvarsigma="7126
\mathchardef\bbvarphi="7127

\def\IL{\relax{\rm I\kern-.18em L}}
\def\IH{\relax{\rm I\kern-.18em H}}
\def\IR{\relax{\rm I\kern-.18em R}}
\def\IC{\relax{\rm I\kern-0.54 em C}}





\def\CA{{\cal A}}

\def\CF{{\cal F}}

\def\CN{{\cal N}}
\def\CO{{\cal O}}

\def\CR{{\cal R}}
\def\CS{{\cal S}}


\def\1{{\ds 1}}
\def\R{\hbox{$\bb R$}}


\noblackbox

\def\unit{\relax{\rm 1\kern-.26em I}}
\def\nada{\relax{\rm 0\kern-.30em l}}
\def\tilde{\widetilde}
\def\t{\tilde}


\noblackbox
\def\IL{\relax{\rm I\kern-.18em L}}
\def\IH{\relax{\rm I\kern-.18em H}}
\def\IR{\relax{\rm I\kern-.18em R}}
\def\IC{\relax\hbox{$\inbar\kern-.3em{\rm C}$}}
\def\IZ{\relax\ifmmode\mathchoice
{\hbox{\cmss Z\kern-.4em Z}}{\hbox{\cmss Z\kern-.4em Z}} {\lower.9pt\hbox{\cmsss Z\kern-.4em Z}}
{\lower1.2pt\hbox{\cmsss Z\kern-.4em Z}}\else{\cmss Z\kern-.4em Z}\fi}

\def\CN {{\cal N}}
\def\CR {{\cal R}}

\def\CF {{\cal F}}

\def\partialslash{\not{\hbox{\kern-2pt $\partial$}}}

\def\CO {{\cal O}}

\def\CS {{\cal S}}
\def\CA{{\cal A}}


\def\CN {{\cal N}}

\def\CO {{\cal O}}

\def\CS {{\cal S }}

\font\manual=manfnt \def\dbend{\lower3.5pt\hbox{\manual\char127}}

\def\IZ{\relax\ifmmode\mathchoice
{\hbox{\cmss Z\kern-.4em Z}}{\hbox{\cmss Z\kern-.4em Z}} {\lower.9pt\hbox{\cmsss Z\kern-.4em Z}}
{\lower1.2pt\hbox{\cmsss Z\kern-.4em Z}}\else{\cmss Z\kern-.4em Z}\fi}
\def\half {{1\over 2}}

\def\bar{\overline}
\def\CS{{\cal S}}

\def\rt2{\sqrt{2}}
\def\irt2{{1\over\sqrt{2}}}

\def\t{\tilde}
\def\hat{\widehat}
\def\slashchar#1{\setbox0=\hbox{$#1$}           
   \dimen0=\wd0                                 
   \setbox1=\hbox{/} \dimen1=\wd1               
   \ifdim\dimen0>\dimen1                        
      \rlap{\hbox to \dimen0{\hfil/\hfil}}      
      #1                                        
   \else                                        
      \rlap{\hbox to \dimen1{\hfil$#1$\hfil}}   
      /                                         
   \fi}

\def\foursqr#1#2{{\vcenter{\vbox{
    \hrule height.#2pt
    \hbox{\vrule width.#2pt height#1pt \kern#1pt
    \vrule width.#2pt}
    \hrule height.#2pt
    \hrule height.#2pt
    \hbox{\vrule width.#2pt height#1pt \kern#1pt
    \vrule width.#2pt}
    \hrule height.#2pt
        \hrule height.#2pt
    \hbox{\vrule width.#2pt height#1pt \kern#1pt
    \vrule width.#2pt}
    \hrule height.#2pt
        \hrule height.#2pt
    \hbox{\vrule width.#2pt height#1pt \kern#1pt
    \vrule width.#2pt}
    \hrule height.#2pt}}}}
\def\psqr#1#2{{\vcenter{\vbox{\hrule height.#2pt
    \hbox{\vrule width.#2pt height#1pt \kern#1pt
    \vrule width.#2pt}
    \hrule height.#2pt \hrule height.#2pt
    \hbox{\vrule width.#2pt height#1pt \kern#1pt
    \vrule width.#2pt}
    \hrule height.#2pt}}}}
\def\sqr#1#2{{\vcenter{\vbox{\hrule height.#2pt
    \hbox{\vrule width.#2pt height#1pt \kern#1pt
    \vrule width.#2pt}
    \hrule height.#2pt}}}}

\def\figin{\epsfcheck\figin}\def\figins{\epsfcheck\figins}
\def\epsfcheck{\ifx\epsfbox\UnDeFiNeD
\message{(NO epsf.tex, FIGURES WILL BE IGNORED)}
\gdef\figin##1{\vskip2in}\gdef\figins##1{\hskip.5in}
\else\message{(FIGURES WILL BE INCLUDED)}%
\gdef\figin##1{##1}\gdef\figins##1{##1}\fi}
\def\DefWarn#1{}
\def\figinsert{\goodbreak\midinsert}
\def\ifig#1#2#3{\DefWarn#1\xdef#1{fig.~\the\figno}
\writedef{#1\leftbracket fig.\noexpand~\the\figno}%
\figinsert\figin{\centerline{#3}}\medskip\centerline{\vbox{\baselineskip12pt \advance\hsize by
-1truein\noindent\footnotefont{\bf Fig.~\the\figno:\ } \it#2}}
\bigskip\endinsert\global\advance\figno by1}


\lref\WessCP{
  J.~Wess and J.~Bagger,
  ``Supersymmetry and supergravity,''
Princeton, USA: Univ. Pr. (1992) 259 p.
}

\lref\KetovES{
  S.~V.~Ketov,
  ``2-d, N=2 and N=4 supergravity and the Liouville theory in superspace,''
Phys.\ Lett.\ B {\bf 377}, 48 (1996).
[hep-th/9602038].
}
\lref\ClossetPDA{
  C.~Closset and S.~Cremonesi,
  ``Comments on N=(2,2) Supersymmetry on Two-Manifolds,''
[arXiv:1404.2636 [hep-th]].
}

\lref\FestucciaWS{
  G.~Festuccia and N.~Seiberg,
  ``Rigid Supersymmetric Theories in Curved Superspace,''
JHEP {\bf 1106}, 114 (2011).
[arXiv:1105.0689 [hep-th]].
}

\lref\MartelliFU{
  D.~Martelli, A.~Passias and J.~Sparks,
  ``The Gravity dual of supersymmetric gauge theories on a squashed three-sphere,''
Nucl.\ Phys.\ B {\bf 864}, 840 (2012).
[arXiv:1110.6400 [hep-th]].
}

\lref\ClossetVG{
  C.~Closset, T.~T.~Dumitrescu, G.~Festuccia, Z.~Komargodski and N.~Seiberg,
  ``Contact Terms, Unitarity, and F-Maximization in Three-Dimensional Superconformal Theories,''
JHEP {\bf 1210}, 053 (2012).
[arXiv:1205.4142 [hep-th]].
}

\lref\ClossetVP{
  C.~Closset, T.~T.~Dumitrescu, G.~Festuccia, Z.~Komargodski and N.~Seiberg,
  ``Comments on Chern-Simons Contact Terms in Three Dimensions,''
JHEP {\bf 1209}, 091 (2012).
[arXiv:1206.5218 [hep-th]].
}

\lref\KomargodskiRB{
  Z.~Komargodski and N.~Seiberg,
  ``Comments on Supercurrent Multiplets, Supersymmetric Field Theories and Supergravity,''
JHEP {\bf 1007}, 017 (2010).
[arXiv:1002.2228 [hep-th]].
}

\lref\GatesHY{
  S.~J.~Gates, Jr.,
   ``Ectoplasm has no topology,''
Nucl.\ Phys.\ B {\bf 541}, 615 (1999).
[hep-th/9809056].
}

\lref\SohniusTP{
  M.~F.~Sohnius and P.~C.~West,
  ``An Alternative Minimal Off-Shell Version of N=1 Supergravity,''
Phys.\ Lett.\ B {\bf 105}, 353 (1981).
}

\lref\DolanRP{
  F.~A.~H.~Dolan, V.~P.~Spiridonov and G.~S.~Vartanov,
  ``From 4d superconformal indices to 3d partition functions,''
Phys.\ Lett.\ B {\bf 704}, 234 (2011).
[arXiv:1104.1787 [hep-th]].
}

\lref\DeserYX{
  S.~Deser and A.~Schwimmer,
  ``Geometric classification of conformal anomalies in arbitrary dimensions,''
Phys.\ Lett.\ B {\bf 309}, 279 (1993).
[hep-th/9302047].
}

\lref\GaddeIA{
  A.~Gadde and W.~Yan,
  ``Reducing the 4d Index to the $S^3$ Partition Function,''
JHEP {\bf 1212}, 003 (2012).
[arXiv:1104.2592 [hep-th]].
}
\lref\FreedmanZZ{
  D.~Z.~Freedman and A.~Van Proeyen,
   ``Supergravity,''
Cambridge, UK: Cambridge Univ. Pr. (2012) 607 p.
}

\lref\deRooMM{
  M.~de Roo, J.~W.~van Holten, B.~de Wit and A.~Van Proeyen,
  ``Chiral Superfields in $N=2$ Supergravity,''
Nucl.\ Phys.\ B {\bf 173}, 175 (1980)..
}

\lref\ImamuraUW{
  Y.~Imamura,
  ``Relation between the 4d superconformal index and the $S^3$ partition function,''
JHEP {\bf 1109}, 133 (2011).
[arXiv:1104.4482 [hep-th]].
}

\lref\DumitrescuHA{
  T.~T.~Dumitrescu, G.~Festuccia and N.~Seiberg,
  ``Exploring Curved Superspace,''
JHEP {\bf 1208}, 141 (2012).
[arXiv:1205.1115 [hep-th]].
}

\lref\DumitrescuAT{
  T.~T.~Dumitrescu and G.~Festuccia,
  ``Exploring Curved Superspace (II),''
JHEP {\bf 1301}, 072 (2013).
[arXiv:1209.5408 [hep-th]].
}

\lref\DumitrescuIU{
  T.~T.~Dumitrescu and N.~Seiberg,
  ``Supercurrents and Brane Currents in Diverse Dimensions,''
JHEP {\bf 1107}, 095 (2011).
[arXiv:1106.0031 [hep-th]].
}

\lref\CasiniKV{
  H.~Casini, M.~Huerta and R.~C.~Myers,
JHEP {\bf 1105}, 036 (2011).
[arXiv:1102.0440 [hep-th]].
}
\lref\SohniusFW{
  M.~Sohnius and P.~C.~West,
  ``The Tensor Calculus And Matter Coupling Of The Alternative Minimal Auxiliary Field Formulation Of N=1 Supergravity,''
Nucl.\ Phys.\ B {\bf 198}, 493 (1982).
}

\lref\KuzenkoXG{
  S.~M.~Kuzenko, U.~Lindstrom and G.~Tartaglino-Mazzucchelli,
  ``Off-shell supergravity-matter couplings in three dimensions,''
JHEP {\bf 1103}, 120 (2011).
[arXiv:1101.4013 [hep-th]].
}

\lref\KuzenkoBC{
  S.~M.~Kuzenko, U.~Lindstrom and G.~Tartaglino-Mazzucchelli,
  ``Three-dimensional (p,q) AdS superspaces and matter couplings,''
JHEP {\bf 1208}, 024 (2012).
[arXiv:1205.4622 [hep-th]].
}

\lref\WittenEV{
  E.~Witten,
  ``Supersymmetric Yang-Mills theory on a four manifold,''
J.\ Math.\ Phys.\  {\bf 35}, 5101 (1994).
[hep-th/9403195].
}

\lref\KlareGN{
  C.~Klare, A.~Tomasiello and A.~Zaffaroni,
  ``Supersymmetry on Curved Spaces and Holography,''
JHEP {\bf 1208}, 061 (2012).
[arXiv:1205.1062 [hep-th]].
}

\lref\KomargodskiPC{
  Z.~Komargodski and N.~Seiberg,
  ``Comments on the Fayet-Iliopoulos Term in Field Theory and Supergravity,''
JHEP {\bf 0906}, 007 (2009).
[arXiv:0904.1159 [hep-th]].
}

\lref\ImamuraSU{
  Y.~Imamura and S.~Yokoyama,
  ``Index for three dimensional superconformal field theories with general R-charge assignments,''
JHEP {\bf 1104}, 007 (2011).
[arXiv:1101.0557 [hep-th]].
}

\lref\OkudaKE{
  T.~Okuda and V.~Pestun,
  ``On the instantons and the hypermultiplet mass of N=2* super Yang-Mills on $S^{4}$,''
JHEP {\bf 1203}, 017 (2012).
[arXiv:1004.1222 [hep-th]].
}

\lref\HoweZM{
  P.~S.~Howe, J.~M.~Izquierdo, G.~Papadopoulos and P.~K.~Townsend,
  ``New supergravities with central charges and Killing spinors in (2+1)-dimensions,''
Nucl.\ Phys.\ B {\bf 467}, 183 (1996).
[hep-th/9505032].
}

\lref\PestunRZ{
  V.~Pestun,
  ``Localization of gauge theory on a four-sphere and supersymmetric Wilson loops,''
Commun.\ Math.\ Phys.\  {\bf 313}, 71 (2012).
[arXiv:0712.2824 [hep-th]].
}

\lref\KuzenkoRD{
  S.~M.~Kuzenko and G.~Tartaglino-Mazzucchelli,
  ``Three-dimensional N=2 (AdS) supergravity and associated supercurrents,''
JHEP {\bf 1112}, 052 (2011).
[arXiv:1109.0496 [hep-th]].
}

\lref\RomelsbergerEC{
  C.~Romelsberger,
  ``Calculating the Superconformal Index and Seiberg Duality,''
[arXiv:0707.3702 [hep-th]].
}

\lref\Paris{
  A.~Romelsberger,
  ``Calculating the Superconformal Index and Seiberg Duality,''
[arXiv:0707.3702 [hep-th]].
http://itf.fys.kuleuven.be/~toine/LectParis.pdf
}

\lref\RomelsbergerEG{
  C.~Romelsberger,
  ``Counting chiral primaries in N = 1, d=4 superconformal field theories,''
Nucl.\ Phys.\ B {\bf 747}, 329 (2006).
[hep-th/0510060].
}

\lref\DolanQI{
  F.~A.~Dolan and H.~Osborn,
  ``Applications of the Superconformal Index for Protected Operators and q-Hypergeometric Identities to N=1 Dual Theories,''
Nucl.\ Phys.\ B {\bf 818}, 137 (2009).
[arXiv:0801.4947 [hep-th]].
}

\lref\ClossetRU{
  C.~Closset, T.~T.~Dumitrescu, G.~Festuccia and Z.~Komargodski,
  ``Supersymmetric Field Theories on Three-Manifolds,''
JHEP {\bf 1305}, 017 (2013).
[arXiv:1212.3388].
}

\lref\KSii{
  K.~Kodaira and D.C.~Spencer,
  ``On Deformations of Complex Analytic Structures II,''
Ann. Math. {\bf 67}, 403 (1958).
}

\lref\Kodairabook{
K.~Kodaira, ``Complex Manifolds and Deformation of Complex Structures,'' Springer (1986).
}

\lref\Kobayashi{
S.~Kobayashi, ``Differential Geometry of Complex Vector Bundles,'' Princeton University Press (1987).
}

\lref\WittenHF{
  E.~Witten,
  ``Quantum Field Theory and the Jones Polynomial,''
Commun.\ Math.\ Phys.\  {\bf 121}, 351 (1989).
}

\lref\WittenDF{
  E.~Witten,
  ``Constraints on Supersymmetry Breaking,''
Nucl.\ Phys.\ B {\bf 202}, 253 (1982)..
}

\lref\spivak{
M.~Spivak, ``Calculus on Manifolds,'' Perseus (1965).
}

\lref\ImamuraWG{
  Y.~Imamura and D.~Yokoyama,
  ``N=2 supersymmetric theories on squashed three-sphere,''
Phys.\ Rev.\ D {\bf 85}, 025015 (2012).
[arXiv:1109.4734 [hep-th]].
}

\lref\MartelliAQA{
  D.~Martelli and A.~Passias,
  ``The gravity dual of supersymmetric gauge theories on a two-parameter deformed three-sphere,''
[arXiv:1306.3893 [hep-th]].
}

\lref\AharonyDHA{
  O.~Aharony, S.~S.~Razamat, N.~Seiberg and B.~Willett,
  ``3d dualities from 4d dualities,''
JHEP {\bf 1307}, 149 (2013).
[arXiv:1305.3924 [hep-th]].
}

\lref\SpiridonovZA{
  V.~P.~Spiridonov and G.~S.~Vartanov,
  ``Elliptic Hypergeometry of Supersymmetric Dualities,''
Commun.\ Math.\ Phys.\  {\bf 304}, 797 (2011).
[arXiv:0910.5944 [hep-th]].
}

\lref\KinneyEJ{
  J.~Kinney, J.~M.~Maldacena, S.~Minwalla and S.~Raju,
  ``An Index for 4 dimensional super conformal theories,''
Commun.\ Math.\ Phys.\  {\bf 275}, 209 (2007).
[hep-th/0510251].
}

\lref\SenPH{
  D.~Sen,
  ``Supersymmetry In The Space-time $\R \times S^3$,''
Nucl.\ Phys.\ B {\bf 284}, 201 (1987).
}

\lref\Kodairasone{
  K.~Kodaira, ``Complex structures on $S^{1}\times S^{3}$,''
Proceedings of the National Academy of Sciences of the United States of America {\bf 55}, 240 (1966).
}

\lref\Belgun{
F.~A.~Belgun, ``On the metric structure of non-K\"ahler complex surfaces,''
Math. Ann. {\bf 317}, 1 (2000).
}

\lref\gaudorn{
P.~Gauduchon and~L.~Ornea, ``Locally conformally K\"ahler metrics on Hopf surfaces,'' Ann. Inst. Fourier {\bf 48}, 4 (1998).
}

\lref\AharonyHDA{
  O.~Aharony, N.~Seiberg and Y.~Tachikawa,
  ``Reading between the lines of four-dimensional gauge theories,''
JHEP {\bf 1308}, 115 (2013).
[arXiv:1305.0318 [hep-th]].
}

\lref\CassaniDBA{
  D.~Cassani and D.~Martelli,
  ``Supersymmetry on curved spaces and superconformal anomalies,''
[arXiv:1307.6567 [hep-th]].
}

\lref\JohansenAW{
  A.~Johansen,
  ``Twisting of $N=1$ SUSY gauge theories and heterotic topological theories,''
Int.\ J.\ Mod.\ Phys.\ A {\bf 10}, 4325 (1995).
[hep-th/9403017].
}

\lref\Spiridonov{
  V.~Spiridonov,
  ``Elliptic Hypergeometric Functions,''
[arXiv:0704.3099].
}

\lref\Rains{
  E.~M.~Rains,
  ``Transformations of Elliptic Hypergeometric Integrals,''
[math/0309252].
}

\lref\Bult{
  F.~van de Bult,
  ``Hyperbolic Hypergeometric Functions,''
University of Amsterdam, Ph.D. Thesis, [http://www.its.caltech.edu/$\sim$vdbult/Thesis.pdf].
}

\lref\KodairaCCASii{
K.~Kodaira, ``On the structure of compact complex analytic surfaces, II,'' American Journal of Mathematics {\bf 88}, 682 (1966).
}

\lref\SpiridonovHF{
  V.~P.~Spiridonov and G.~S.~Vartanov,
  ``Elliptic hypergeometry of supersymmetric dualities II. Orthogonal groups, knots, and vortices,''
[arXiv:1107.5788 [hep-th]].
}

\lref\BeniniNC{
  F.~Benini, T.~Nishioka and M.~Yamazaki,
  ``4d Index to 3d Index and 2d TQFT,''
Phys.\ Rev.\ D {\bf 86}, 065015 (2012).
[arXiv:1109.0283 [hep-th]].
}

\lref\RazamatOPA{
  S.~S.~Razamat and B.~Willett,
  ``Global Properties of Supersymmetric Theories and the Lens Space,''
[arXiv:1307.4381 [hep-th]].
}

\lref\Nakagawa{
N.~Nakagawa, ``Complex structures on $L (p, q)\times S^1$,'' Hiroshima Mathematical Journal {\bf 25} 423 (1995).
}

\lref\GaddeDDA{
  A.~Gadde and S.~Gukov,
  ``2d Index and Surface operators,''
[arXiv:1305.0266 [hep-th]].
}

\lref\BeniniNDA{
  F.~Benini, R.~Eager, K.~Hori and Y.~Tachikawa,
  ``Elliptic genera of two-dimensional N=2 gauge theories with rank-one gauge groups,''
[arXiv:1305.0533 [hep-th]].
}

\lref\BeniniXPA{
  F.~Benini, R.~Eager, K.~Hori and Y.~Tachikawa,
  ``Elliptic genera of 2d N=2 gauge theories,''
[arXiv:1308.4896 [hep-th]].
}

\lref\KapustinKZ{
  A.~Kapustin, B.~Willett and I.~Yaakov,
  ``Exact Results for Wilson Loops in Superconformal Chern-Simons Theories with
  Matter,''
  JHEP {\bf 1003}, 089 (2010)
  [arXiv:0909.4559 [hep-th]].
}

\lref\JafferisZI{
  D.~L.~Jafferis, I.~R.~Klebanov, S.~S.~Pufu and B.~R.~Safdi,
 ``Towards the F-Theorem: N=2 Field Theories on the Three-Sphere,''
JHEP {\bf 1106}, 102 (2011).
[arXiv:1103.1181 [hep-th]].
}
\lref\JafferisUN{
  D.~L.~Jafferis,
  ``The Exact Superconformal R-Symmetry Extremizes Z,''
JHEP {\bf 1205}, 159 (2012).
[arXiv:1012.3210 [hep-th]].
}

\lref\CecottiSA{
  S.~Cecotti,
  ``Higher Derivative Supergravity Is Equivalent To Standard Supergravity Coupled To Matter. 1.,''
Phys.\ Lett.\ B {\bf 190}, 86 (1987)..
}

\lref\HamaAV{
  N.~Hama, K.~Hosomichi and S.~Lee,
  ``Notes on SUSY Gauge Theories on Three-Sphere,''
JHEP {\bf 1103}, 127 (2011).
[arXiv:1012.3512 [hep-th]].
}
\lref\ParkNN{
  D.~S.~Park and J.~Song,
  ``The Seiberg-Witten Kahler Potential as a Two-Sphere Partition Function,''
JHEP {\bf 1301}, 142 (2013).
[arXiv:1211.0019 [hep-th]].
}

\lref\HamaEA{
  N.~Hama, K.~Hosomichi and S.~Lee,
  ``SUSY Gauge Theories on Squashed Three-Spheres,''
JHEP {\bf 1105}, 014 (2011).
[arXiv:1102.4716 [hep-th]].
}

\lref\AldayLBA{
  L.~F.~Alday, D.~Martelli, P.~Richmond and J.~Sparks,
  ``Localization on Three-Manifolds,''
[arXiv:1307.6848 [hep-th]].
}

\lref\NianQWA{
  J.~Nian,
  ``Localization of Supersymmetric Chern-Simons-Matter Theory on a Squashed $S^3$ with $SU(2)\times U(1)$ Isometry,''
[arXiv:1309.3266 [hep-th]].
}
\lref\JockersDK{
  H.~Jockers, V.~Kumar, J.~M.~Lapan, D.~R.~Morrison and M.~Romo,
  ``Two-Sphere Partition Functions and Gromov-Witten Invariants,''
Commun.\ Math.\ Phys.\  {\bf 325}, 1139 (2014).
[arXiv:1208.6244 [hep-th]].
}

\lref\KlareDKA{
  C.~Klare and A.~Zaffaroni,
  ``Extended Supersymmetry on Curved Spaces,''
JHEP {\bf 1310}, 218 (2013).
[arXiv:1308.1102 [hep-th]].
}

\lref\Toine{
  A. Van Proeyen
  ``${\cal N}=2$ Supergravity in $d=4,5,6$ and its Matter Couplings.''
http://itf.fys.kuleuven.be/$\sim$toine/LectParis.pdf
}

\lref\BhattacharyaZY{
  J.~Bhattacharya, S.~Bhattacharyya, S.~Minwalla and S.~Raju,
  ``Indices for Superconformal Field Theories in 3,5 and 6 Dimensions,''
JHEP {\bf 0802}, 064 (2008).
[arXiv:0801.1435 [hep-th]].
}

\lref\KapustinJM{
  A.~Kapustin and B.~Willett,
  ``Generalized Superconformal Index for Three Dimensional Field Theories,''
[arXiv:1106.2484 [hep-th]].
}

\lref\DimoftePY{
  T.~Dimofte, D.~Gaiotto and S.~Gukov,
  ``3-Manifolds and 3d Indices,''
[arXiv:1112.5179 [hep-th]].
}

\lref\BeniniUI{
  F.~Benini and S.~Cremonesi,
  ``Partition functions of N=(2,2) gauge theories on $S^2$ and vortices,''
[arXiv:1206.2356 [hep-th]].
}

\lref\DoroudXW{
  N.~Doroud, J.~Gomis, B.~Le Floch and S.~Lee,
  ``Exact Results in D=2 Supersymmetric Gauge Theories,''
JHEP {\bf 1305}, 093 (2013).
[arXiv:1206.2606 [hep-th]].
}

\lref\DoroudPKA{
  N.~Doroud and J.~Gomis,
  ``Gauge Theory Dynamics and Kahler Potential for Calabi-Yau Complex Moduli,''
[arXiv:1309.2305 [hep-th]].
}

\lref\CecottiQE{
  S.~Cecotti, S.~Ferrara, M.~Porrati and S.~Sabharwal,
  ``New Minimal Higher Derivative Supergravity Coupled To Matter,''
Nucl.\ Phys.\ B {\bf 306}, 160 (1988)..
}

\lref\KlareGN{
  C.~Klare, A.~Tomasiello and A.~Zaffaroni,
  ``Supersymmetry on Curved Spaces and Holography,''
JHEP {\bf 1208}, 061 (2012).
[arXiv:1205.1062 [hep-th]].
}

\lref\BonelliMMA{
  G.~Bonelli, A.~Sciarappa, A.~Tanzini and P.~Vasko,
  ``Vortex partition functions, wall crossing and equivariant Gromov-Witten invariants,''
[arXiv:1307.5997 [hep-th]].
}

\lref\GreenDA{
  D.~Green, Z.~Komargodski, N.~Seiberg, Y.~Tachikawa and B.~Wecht,
  ``Exactly Marginal Deformations and Global Symmetries,''
JHEP {\bf 1006}, 106 (2010).
[arXiv:1005.3546 [hep-th]].
}

\lref\KutasovXB{
  D.~Kutasov,
  ``Geometry On The Space Of Conformal Field Theories And Contact Terms,''
Phys.\ Lett.\ B {\bf 220}, 153 (1989).
}

\lref\LeighEP{
  R.~G.~Leigh and M.~J.~Strassler,
  ``Exactly marginal operators and duality in four-dimensional N=1 supersymmetric gauge theory,''
Nucl.\ Phys.\ B {\bf 447}, 95 (1995).
[hep-th/9503121].
}

\lref\ZamolodchikovGT{
  A.~B.~Zamolodchikov,
  ``Irreversibility of the Flux of the Renormalization Group in a 2D Field Theory,''
JETP Lett.\  {\bf 43}, 730 (1986), [Pisma Zh.\ Eksp.\ Teor.\ Fiz.\  {\bf 43}, 565 (1986)].
}

\lref\SeibergPF{
  N.~Seiberg,
  ``Observations on the Moduli Space of Superconformal Field Theories,''
Nucl.\ Phys.\ B {\bf 303}, 286 (1988).
}

\lref\CecottiME{
  S.~Cecotti and C.~Vafa,
  ``Topological antitopological fusion,''
Nucl.\ Phys.\ B {\bf 367}, 359 (1991).
}

\lref\NishiokaHAA{
  T.~Nishioka and I.~Yaakov,
  ``Supersymmetric Renyi Entropy,''
[arXiv:1306.2958 [hep-th]].
}

\lref\ClossetVRA{
  C.~Closset, T.~T.~Dumitrescu, G.~Festuccia and Z.~Komargodski,
  ``The Geometry of Supersymmetric Partition Functions,''
[arXiv:1309.5876 [hep-th]].
}

\lref\deWitZA{
  B.~de Wit, S.~Katmadas and M.~van Zalk,
  ``New supersymmetric higher-derivative couplings: Full N=2 superspace does not count!,''
JHEP {\bf 1101}, 007 (2011).
[arXiv:1010.2150 [hep-th]].
}

\lref\Hijazi{
O. Hijazi, 
``A conformal lower bound for the smallest eigenvalue of the Dirac operator and Killing spinors,'' Communications in Mathematical Physics 104 (1986), no. 1, 151Ð162.
}

\lref\deRooMM{
  M.~de Roo, J.~W.~van Holten, B.~de Wit and A.~Van Proeyen,
  ``Chiral Superfields in $N=2$ Supergravity,''
Nucl.\ Phys.\ B {\bf 173}, 175 (1980).
}

\lref\OsbornGM{
  H.~Osborn,
  ``Weyl consistency conditions and a local renormalization group equation for general renormalizable field theories,''
Nucl.\ Phys.\ B {\bf 363}, 486 (1991).
}

\lref\DeserYX{
  S.~Deser and A.~Schwimmer,
  ``Geometric classification of conformal anomalies in arbitrary dimensions,''
Phys.\ Lett.\ B {\bf 309}, 279 (1993).
[hep-th/9302047].
}

\lref\Huybrechts{
D.~Huybrechts, ``Complex Geometry: An Introduction,'' Springer (2006).
}

\lref\BizetUUA{
  N.~C.~Bizet, A.~Klemm and D.~V.~Lopes,
  ``Landscaping with fluxes and the E8 Yukawa Point in F-theory,''
[arXiv:1404.7645 [hep-th]].
}
\lref\TanakaDCA{
  A.~Tanaka,
  ``Localization on round sphere revisited,''
[arXiv:1309.4992 [hep-th]].
}

\lref\AsninXX{
  V.~Asnin,
  ``On metric geometry of conformal moduli spaces of four-dimensional superconformal theories,''
JHEP {\bf 1009}, 012 (2010).
[arXiv:0912.2529 [hep-th]].
}

\lref\DoroudXW{
  N.~Doroud, J.~Gomis, B.~Le Floch and S.~Lee,
  ``Exact Results in $D=2$ Supersymmetric Gauge Theories,''
JHEP {\bf 1305}, 093 (2013).
[arXiv:1206.2606 [hep-th]].
}

\lref\CasiniEI{
  H.~Casini and M.~Huerta,
  ``On the RG running of the entanglement entropy of a circle,''
Phys.\ Rev.\ D {\bf 85}, 125016 (2012).
[arXiv:1202.5650 [hep-th]].
}

\lref\LercheUY{
  W.~Lerche, C.~Vafa and N.~P.~Warner,
  ``Chiral Rings in N=2 Superconformal Theories,''
Nucl.\ Phys.\ B {\bf 324}, 427 (1989).
}

\lref\GomisWY{
  J.~Gomis and S.~Lee,
  ``Exact K\"ahler Potential from Gauge Theory and Mirror Symmetry,''
JHEP {\bf 1304}, 019 (2013).
[arXiv:1210.6022 [hep-th]].
}

\lref\ClossetVP{
  C.~Closset, T.~T.~Dumitrescu, G.~Festuccia, Z.~Komargodski and N.~Seiberg,
  ``Comments on Chern-Simons Contact Terms in Three Dimensions,''
JHEP {\bf 1209}, 091 (2012).
[arXiv:1206.5218 [hep-th]].
}

\lref\BeniniUI{
  F.~Benini and S.~Cremonesi,
  ``Partition functions of ${\cal N}=(2,2)$ gauge theories on $S^2$ and vortices,''
[arXiv:1206.2356 [hep-th]].
}

\lref\ClossetSXA{
  C.~Closset and I.~Shamir,
  ``The $\CN=1$ Chiral Multiplet on $T^2\times S^2$ and Supersymmetric Localization,''
[arXiv:1311.2430 [hep-th]].
}

\lref\BeniniMF{
  F.~Benini, C.~Closset and S.~Cremonesi,
  ``Comments on 3d Seiberg-like dualities,''
JHEP {\bf 1110}, 075 (2011).
[arXiv:1108.5373 [hep-th]].
}

\lref\IseHopfsurf{
M.~Ise,
   ``On the geometry of Hopf manifolds,''
  Osaka Math. J., 12, 1960, p.387--402.
}

\lref\CardyCWA{
  J.~L.~Cardy,
  ``Is There a c Theorem in Four-Dimensions?,''
Phys.\ Lett.\ B {\bf 215}, 749 (1988).
}

\lref\JackEB{
  I.~Jack and H.~Osborn,
  ``Analogs for the $c$ Theorem for Four-dimensional Renormalizable Field Theories,''
Nucl.\ Phys.\ B {\bf 343}, 647 (1990).
}

\lref\KomargodskiVJ{
  Z.~Komargodski and A.~Schwimmer,
  ``On Renormalization Group Flows in Four Dimensions,''
JHEP {\bf 1112}, 099 (2011).
[arXiv:1107.3987 [hep-th]].
}

\lref\PapadodimasEU{
  K.~Papadodimas,
  ``Topological Anti-Topological Fusion in Four-Dimensional Superconformal Field Theories,''
JHEP {\bf 1008}, 118 (2010).
[arXiv:0910.4963 [hep-th]].
}

\lref\KomargodskiXV{
  Z.~Komargodski,
  ``The Constraints of Conformal Symmetry on RG Flows,''
JHEP {\bf 1207}, 069 (2012).
[arXiv:1112.4538 [hep-th]].
}

\lref\MallHopfsurf{
D.~Mall,
   ``The cohomology of line bundles on Hopf manifolds,''
  Osaka J. Math., 28, 1991, p.999--1015.
}

\lref\StromingerPD{
  A.~Strominger,
  ``Special Geometry,''
Commun.\ Math.\ Phys.\  {\bf 133}, 163 (1990).
}

\lref\BardeenPM{
  W.~A.~Bardeen and B.~Zumino,
  ``Consistent and Covariant Anomalies in Gauge and Gravitational Theories,''
Nucl.\ Phys.\ B {\bf 244}, 421 (1984).
}

\lref\CallanSA{
  C.~G.~Callan, Jr. and J.~A.~Harvey,
  ``Anomalies and Fermion Zero Modes on Strings and Domain Walls,''
Nucl.\ Phys.\ B {\bf 250}, 427 (1985).
}

\lref\CecottiME{
  S.~Cecotti and C.~Vafa,
  ``Topological antitopological fusion,''
Nucl.\ Phys.\ B {\bf 367}, 359 (1991).
}

\lref\AsninXX{
  V.~Asnin,
  ``On metric geometry of conformal moduli spaces of four-dimensional superconformal theories,''
JHEP {\bf 1009}, 012 (2010).
[arXiv:0912.2529 [hep-th]].
}

\lref\deWitZA{
  B.~de Wit, S.~Katmadas and M.~van Zalk,
  ``New supersymmetric higher-derivative couplings: Full N=2 superspace does not count!,''
JHEP {\bf 1101}, 007 (2011).
[arXiv:1010.2150 [hep-th]].
}

\lref\ButterLTA{
  D.~Butter, B.~de Wit, S.~M.~Kuzenko and I.~Lodato,
  ``New higher-derivative invariants in N=2 supergravity and the Gauss-Bonnet term,''
JHEP {\bf 1312}, 062 (2013).
[arXiv:1307.6546 [hep-th], arXiv:1307.6546].
}

\lref\CecottiQN{
  S.~Cecotti, S.~Ferrara and L.~Girardello,
  ``Geometry of Type II Superstrings and the Moduli of Superconformal Field Theories,''
Int.\ J.\ Mod.\ Phys.\ A {\bf 4}, 2475 (1989).
}



\rightline{NSF-KITP-14-051}
\rightline{WIS/04/14-MAY-DPPA}
\vskip-45pt
\Title{
} {\vbox{\centerline{Sphere Partition Functions and the Zamolodchikov Metric }
}}
\centerline{Efrat Gerchkovitz,$^1$ Jaume Gomis,$^{2}$ and Zohar Komargodski$^1$}
\vskip15pt

\centerline{ $^{1}$ {\it Weizmann Institute of Science, Rehovot
76100, Israel}}
\centerline{$^{2}$ {\it Perimeter Institute for Theoretical Physics, Waterloo, Ontario, N2L 2Y5, Canada}}

\vskip20pt

\centerline{\bf Abstract}
\noindent   

We study the finite part of the sphere partition function of $d$-dimensional Conformal Field Theories (CFTs) as a function of exactly marginal couplings. In odd dimensions, this quantity is physical and independent of the exactly marginal couplings. 
In even dimensions, this object is generally regularization scheme dependent and thus unphysical.
However, in the presence of additional symmetries, the partition function of even-dimensional CFTs can become physical.  For two-dimensional~$\CN=(2,2)$ supersymmetric CFTs, the continuum partition function exists and computes the K\"ahler potential on the chiral and twisted chiral superconformal manifolds.  We provide a new elementary proof of this   result using Ward identities on the sphere. The K\"ahler transformation ambiguity is identified with a local term in the corresponding $\CN=(2,2)$ supergravity theory.  We derive an analogous, new,  result in the case of four-dimensional  $\CN=2$  supersymmetric CFTs:  the $S^4$ partition function computes the K\"ahler potential on the superconformal manifold. Finally, we show that $\CN=1$ supersymmetry in four dimensions and $\CN=(1,1)$ supersymmetry in two dimensions are not sufficient to make the corresponding sphere partition functions well-defined functions of the exactly marginal parameters. 

\Date{May 2014}

\listtoc \writetoc

\newsec{Conformal Manifolds and Sphere Partition Functions} 

Suppose that in a $d$-dimensional CFT, denoted by $p$, there are exactly marginal operators $\{O_i\}$. This means that  $\hbox{dim}(O_i)=d$. In addition, if we deform the theory $p$ by\foot{We choose a normalization that will make later formulae simpler.}
\eqn\deform{{1\over \pi^{d/2}}\int d^dx \lambda^i O_i(x)\,,} 
then, at least for sufficiently small $\lambda^i$, we get a conformal field theory. 
Thus, in some neighborhood of the reference conformal field theory $p$, there is   a family of CFTs.  We will refer to this family of CFTs as the ``conformal manifold,'' and denote it by $\cal S$.  This space admits a natural Riemannian metric,   the Zamolodchikov metric $g_{ij}(p)$, given by~\ZamolodchikovGT\ 
\eqn\Zamometric{\langle O_i(x) O_j(0)\rangle_{p\in {\cal S}}={g_{ij}(p)\over  x^{2d}}~,}
where $0\neq x\in \IR^{d}$. The couplings $\lambda^i$ in \deform\ are coordinates on the conformal manifold $\cal S$ and the operators $O_i$ represent vector fields in ${\cal S}$. Thus, under a change of variables $\lambda'^i=f^i(\lambda^j)$    the metric $g_{ij}$ transforms as a symmetric tensor. 

Consider an observable ${\cal O}$. Suppose we are interested in computing this observable as a function of $\lambda^i$. Starting from some reference CFT at $\lambda^i=0$, we can thus define
\eqn\correla{
\vev{{\cal O}}_\lambda=\vev{{\cal O} \, \exp\left[{{1\over \pi^{d/2}}\int d^dx \lambda^i O_i(x)}\right]}\,.}
In practice, one can expand in the $\lambda^i$ around the reference CFT,  and define \correla\  in some small neighborhood of $\lambda^i=0$ via this expansion. More explicitly, 
\eqn\definition{ \vev{\CO}_{\lambda}=\sum_k{1\over k!}\biggl\langle\CO\left({1\over \pi^{d/2}}\int d^dx \lambda^iO_i(x)\right)^k\biggr\rangle ~.}

It is very important that  \definition\ is  ambiguous. Indeed, even if we solve the reference theory at $\lambda=0$ completely, i.e.~we know all the correlation functions at separated points, we still need the integrated correlation functions in order to compute~\definition. Integrated correlation functions 
generically have ultraviolet divergences. They need to be regularized  by introducing a cutoff and appropriate counterterms need to be added to the  effective action so that the correlation functions are finite in the continuum limit. Therefore, computations of various observables via~\definition\ may depend on the regularization scheme, and consequently be subject to ambiguities.

Let us now analyze these ambiguities in some detail. The formalism where the couplings  are taken to be dimensionless  functions of space rather than constants $\lambda^i\rightarrow \lambda^i(x)$ allows to easily classify all such ambiguities. 
Different regularization schemes simply differ  by local terms of dimension equal or smaller than $d$ in the effective action. 
In order to preserve Ward identities, one may need to fine-tune some of the counterterms.

Let us briefly describe a few representative examples. The finite counterterm 
\eqn\conne{
\int d^dx\,  \delta\Gamma_{ij}^k(x) \lambda^i(x) \lambda^j(x) O_k(x)\, }
  generates a contact term in the operator product expansion  $O_i(x)O_j(0)\sim \delta\Gamma_{ij}^k\delta^d(x)O_k(0)$, and affects various integrated correlation functions. The object  $\delta \Gamma_{ij}^k$ shifts the connection on the conformal manifold ${\cal S}$ \KutasovXB.
  In even dimensions, the local counterterm 
  \eqn\quadrat{
  \Lambda_{UV}^{d/2}\int d^dx\, C_{ij} \lambda^i(x) \lambda^j(x) L(x)\,,}
where $L$ is a primary operator of $\hbox{dim}(L)=d/2$,  captures the ultraviolet divergence that appears in computing the correlator  $\langle L(0) O_i(y)\rangle_\lambda$ 
in conformal perturbation theory. This term  must be carefully tuned to guarantee that  $\langle L(0) O_i(y)\rangle_\lambda=0$ (for $y\neq 0$), which follows from conformal Ward identities. Indeed, to leading order in the expansion~\definition\ we get  $\int d^d x \langle L(0) O_i(y) O_j(x)\rangle$. This correlator is infrared finite but has an ultraviolet divergence from the region $x\rightarrow y$, where  it behaves as $(x-y)^{-3d/2}$. This   divergence   can be regularized by cutting a little sphere of radius $1/ \Lambda_{UV}$ around $x=y$  and canceled by tuning   the local term \quadrat. Finally, another example of a counterterm is
\eqn\zamoct{
\Lambda_{UV}^d \int d^dx\, C_{ij}  \lambda^i(x) \lambda^j(x)~,}
with $C_{ij}$ any symmetric tensor. This counterterm accounts for a power divergent piece that appears in evaluating $\vev{1}_\lambda$ to second order in conformal perturbation theory. 
 
Our primary interest in this paper lies in CFTs on curved spaces. A conformal field theory can be placed canonically on any conformally-flat manifold, in particular, on $S^d$ (in this case we can use the stereographic map). Since there are no infrared divergences, we can study the partition function $Z_{S^d}$. We can repeat this procedure at any point on the conformal manifold. This provides an interesting probe of the conformal manifold, $Z_{S^d}(p)$.

While infrared divergences are absent, ultraviolet divergences remain. As in our discussion above, they are classified by diffeomorphism invariant local terms of dimension~$\leq d$ constructed from the background fields $\lambda^i$ and the space-time metric $g_{mn}$. One finds the following general answer for the partition function as a function of the point $\lambda^i$ on the conformal manifold:
\eqn\spherepa{\eqalign{&d=2n:\qquad \log Z_{S^{2n}}=A_1(\lambda^i)(r\Lambda_{UV})^{2n}+A_2(\lambda^i)(r\Lambda_{UV})^{2n-2}+...+A_{n}(\lambda^i)(r\Lambda_{UV})^2 \cr &+A(\lambda^i)\log(r\Lambda_{UV})+F_{2n}(\lambda^{i})~.  }}
\eqn\spherepai{\eqalign{&d=2n+1:\qquad \log Z_{S^{2n+1}}=B_1(\lambda^i)(r\Lambda_{UV})^{2n+1}+B_2(\lambda^i)(r\Lambda_{UV})^{2n-1}\cr&
+\ldots+B_{n+1}(\lambda^i)(r\Lambda_{UV})  +F_{2n+1}(\lambda^{i})\,, }}
where $r$ is the radius of $S^d$.
The power-law divergent terms correspond to counterterms of the type 
\eqn\diveven{
\Lambda_{UV}^{2n-2k+2}\int d^{2n}x\sqrt g A_k(\lambda^i) \CR^{k-1}} in even dimensions and 
\eqn\dibodd{
 \Lambda_{UV}^{2n-2k+3}\int d^{2n+1}x\sqrt g B_k(\lambda^i) \CR^{k-1}\,}
 in odd dimensions.
 Therefore, all the power-law  divergent terms in~\spherepa\ and~\spherepai\ can be tuned to zero in the continuum limit.
 
 In even dimensions,  the sphere partition function has a  logarithmic dependence on the radius (see \spherepa), which  
 cannot be canceled by a local counterterm. It is associated to the Weyl anomaly. The variation of the partition function under a Weyl transformation with parameter $\sigma$ contains $\int d^{2n}x\sqrt g\, \sigma A(\lambda^i)E_{2n}$, where $E_{2n}$ is the Euler density (the other terms in the Weyl anomaly vanish on the sphere, see~\DeserYX). However, this violates the Wess-Zumino consistency condition unless $A(\lambda^i)=A$, i.e. $A(\lambda^i)$ is a constant. This is then identified with the usual $A$-type anomaly~\DeserYX, which is therefore independent of exactly marginal deformations.\foot{We thank L. Di Pietro and A.~Schwimmer for discussions. } This is necessary for its interpretation as a monotonic function under renormalization group flows~\refs{\ZamolodchikovGT,\CardyCWA,\JackEB,\KomargodskiVJ,\KomargodskiXV}. 
Finally, the function $F_{2n}(\lambda^i)$ in \spherepa\ is ambiguous, as it can be removed by the local counterterm\foot{Note that this counterterm does not affect the partition function on the cylinder $S^{d-1}\times S^1$. The latter has a natural normalization via radial quantization.}
 \eqn\fincteven{
 \int d^{2n}x\sqrt g F_{2n}(\lambda^i)E_{2n}\,.}

In summary, we have shown  that the only physical data in the continuum limit of  $Z_{S^{2n}}$ is the A-anomaly, which is independent of the exactly marginal parameters.

For odd dimensions, absent additional restrictions on the counterterms, we have seen above that all the $B_i$ are ambiguous and can be tuned to zero  (a logarithmic term is absent because one cannot write an appropriate local anomaly polynomial in odd dimensions.) Importantly, however, there is no counterterm for $F_{2n+1}(\lambda^i)$ in \spherepai. More precisely, the only conceivable dimensionless   counterterm would be a  gravitational Chern-Simons term
\eqn\gravcs{
 \int   C(\lambda)\, \Omega^{(2n+1)}\,,}
but because of coordinate invariance it cannot depend on the $\lambda^i$, i.e. $C(\lambda)=\hbox{constant}$. Moreover, it has to have an imaginary coefficient due to CPT symmetry. 
Hence, the real part of $F_{2n+1}(\lambda^i)$ is an unambiguous physical observable and is calculable in any choice of regularization scheme that preserves coordinate invariance.\foot{The imaginary part is more subtle. Only its fractional part is well defined. See, for example, the discussions in~\refs{\WittenHF,\ClossetVG,\ClossetVP}. We will not comment any further on the imaginary part of $F_{2n+1}$.} It measures the finite entanglement entropy across a $S^{2n-1}$ in $\IR^{2n,1}$~\CasiniKV.\foot{The entanglement entropy provides  another way to see that the finite part in even dimensions is ambiguous while in odd dimensions it is physical. Indeed, it is straightforward to write finite local counterterms on the entangling surface of even-dimensional spheres, while in odd dimensions this is impossible. For example, in $d=3$, the entangling surface is a circle, and the finite counterterm $\int_{S^1}|\kappa| dl$  is forbidden because the absolute value renders it nonlocal, while without the absolute value symbol it is not consistent with the vacuum being a pure state. We thank S.~Pufu for discussing this with us.}

We now show that $F_{2n+1}(\lambda_i)$ is constant on the conformal manifold ${\cal S}$. Start at an arbitrary point on the conformal manifold and expand to second order  
\eqn\expandZ{
\log Z[\lambda]\simeq \log Z(0) +{\lambda^i\over \pi^{d/2}} \int d^dx\,\sqrt{g} \vev{O_i}+   {\lambda^i \lambda^j \over  2 \pi^{d}} \int d^dx\,\sqrt{g} \int d^dy \,\sqrt{g} \vev{O_i(x) O_j(y)}  \,.}
Conformal Ward identities on the sphere imply that
\eqn\onept{
\vev{O_i}=0\,.
}
Since this is true at every point on the conformal manifold, we conclude that the continuum conformal field theory sphere partition function is independent of exactly marginal parameters. As we will see, this argument cannot be repeated in even dimensions because there are finite counterterms. Indeed, we will see examples where the sphere partition function does depend on exactly marginal parameters.

We can understand this simplicity of the sphere partition function in odd dimensions from another point of view. The integrated two-point function of exactly marginal operators in the last term of~\expandZ\ is ultraviolet divergent. The singularity arises from the domain where $x\rightarrow y$. The two-point function on the sphere can be obtained from the corresponding two-point function in flat space by a stereographic map. Hence, the integrated two-point function in~\expandZ\ is proportional to
\eqn\integrated{\hbox{vol}(S^{d})\int_{S^{d}}d^{d}y\sqrt g \left({1\over d(0,y)}\right)^{2d}\,,}
where $d(x,y)$ is the $SO(d+1)$ invariant distance on the sphere $S^{d}$. This can be    regulated  by replacing the power $2d$ in the integrand by $2d-2\epsilon$. We can then evaluate the integral in a region where it converges, analytically continuing to the region of interest at the end, to yield
\eqn\anali{
\hbox{lim}_{\,\epsilon\rightarrow 0} {\hbox{vol}(S^d)\hbox{vol}(S^{d-1})\over 2^{d+1}} {\Gamma(d/2)\Gamma(-d/2+\epsilon)\over \Gamma(\epsilon)}\longrightarrow 0\ \ \  \ \hbox{for}\ d=2n+1\,.}
Therefore the integral vanishes in odd dimensions, and since we have shown that there are no finite counterterms in odd dimensions, the answer computed in  any other choice of regularization scheme   would yield the same result. The fact that the second term in the expansion~\expandZ\ vanishes holds true around any point on the conformal manifold. This is consistent with the independence of the partition function of exactly marginal couplings. Note that the same regularization scheme yields a nonzero answer in even dimensions. This will be important below.

In summary, we have shown  that the (real part of the) finite part of  $Z_{S^{2n+1 }}$  in the continuum limit   is physical and unambiguous, and  is independent of the exactly marginal parameters of the conformal field theory. This is a necessary requirement for the finite part of $Z_{S^{2n+1 }}$  to serve as a candidate monotonic function under renormalization group flows (see \refs{\CasiniKV,\JafferisZI,\CasiniEI}).

\newsec{Sphere Partition Functions of Superconformal Field Theories}

Our analysis of the ambiguities  of the sphere partition function of a CFT  followed from assuming  that the partition function can be regulated in a diffeomorphism invariant way.  We then classified all the diffeomorphism invariant counterterms and determined their influence on the sphere partition function. Starting in this section,   we pursue  an analogous analysis for superconformal field theories (SCFTs). Imposing that the sphere partition function of SCFT's can be regulated in a supersymmetric way, we show that the partition function of SCFTs with various amounts of supersymmetry and in various dimensions   have (not unexpectedly) a more restricted set of ambiguities.   This  makes  the sphere partition function of such SCFTs interesting, rich  observables:    the sphere partition functions  of some of these theories are known to compute the K\"ahler potential on the conformal manifold~\refs{\JockersDK,\GomisWY}.

We now outline the logic of our arguments. Consider a $d$-dimensional SCFT and place it on the sphere by stereographic projection. Such a SCFT is by definition invariant under the corresponding superconformal algebra. As we explained in section~1, however, the partition function of a CFT suffers from ultraviolet divergences that arise when exploring the dependence of the partition function on the conformal manifold ${\cal S}$. We now assume that we can regulate the SCFT partition function while preserving the subalgebra of the superconformal algebra that closes into the super-isometries of the $d$-dimensional supersymmetric sphere, which projects out all conformal generators of $S^d$. This   is the general supersymmetry algebra of a  massive supersymmetric  theory on the $d$-dimensional   sphere.

In the supersymmetric context, the local counterterms that parametrize the ambiguities of the partition function are now {\it supergravity} counterterms. In supersymmetric theories, the coordinates that parametrize the conformal manifold are the bottom component of an apropriate  supersymmetry multiplet. Therefore we must consider the most general dimension $\leq d$ locally supersymmetric couplings of these multiplets to the corresponding supergravity theory. The supergravity theory that must be considered is found by embedding the rigid supersymmetric $d$-dimensional sphere  as a supersymmetric background of $d$-dimensional off-shell supergravity,    in the framework put forward by \FestucciaWS. By constructing marginal supergravity counterterms in a given supergravity theory, we can determine the ambiguities  and the leftover physical content of the partition function of SCFTs on $S^d$.

Let us briefly summarize our main results explained in the rest of the paper. In section~3 we discuss two-dimensional ${\cal N}=(2,2)$ SCFTs. It is   known that the conformal manifold of such theories is a K\"ahler manifold, and locally takes a direct product form ${\cal S}_{\hbox{c}}\times {\cal S}_{\hbox{tc}},$ where ${\cal S}_{\hbox{c}}$, and ${\cal S}_{\hbox{tc}}$ are the  chiral and twisted chiral manifolds, respectively.
These theories can be placed on supersymmetric $S^2$ in two different ways while preserving four supercharges~\refs{\DoroudXW,\BeniniUI,\GomisWY,\DoroudPKA}. For each choice there is a corresponding two-dimensional ${\cal N}=(2,2)$ supergravity theory, which we denote by A and B. Our notation parallels that of~\refs{\DoroudXW,\DoroudPKA},  where the corresponding supersymmetry algebras were labeled by $SU(2|1)_A$ and $SU(2|1)_B$. Imposing that the SCFT partition function preserves $SU(2|1)_B$, we find that there is a finite B-type supergravity counterterm~\KetovES\
\eqn\localtwod{\int d^2\Theta\, \varepsilon^{-1}  R\, \CF(\Phi^i)+c.c.~,}
 where $\Phi^i$ are   chiral superfields, whose bottom components are the coordinates $\lambda^i$ in ${\cal S}_{\hbox{c}}$, $R$ is the supergravity curvature superfield and $\varepsilon^{-1}$ is the supergravity chiral superspace measure. A similar A-type  supergravity counterterm can be constructed for a SCFT preserving $SU(2|1)_A$ by replacing chiral superfields by twisted chiral superfieds $\Omega$, whose bottom components $\lambda^i$ now span ${\cal S}_{\hbox{tc}}$. When these counterterms are evaluated on the supersymmetric $S^2$ background they give rise to the following K\"ahler ambiguities in the partition function 
 \eqn\ambigu{
 Z_{S^2}^A\simeq Z^A_{S^2} \,e^{-\CF(\lambda^i)-\bar \CF(\bar \lambda^{\bar i})}~,\qquad Z_{S^2}^B\simeq Z^B_{S^2}\,e^{-\CF(\lambda^{ i})-\bar \CF(\bar \lambda^{\bar i})}\,.}
 If we describe the same ${\cal N}=(2,2)$ SCFT in two different ways, with different choices of counterterms,  then the partition functions may differ by~\ambigu. This means that the two-sphere partition functions $Z_A$ and $Z_B$ are generally not  functions, but  rather sections: the transitions between different patches may involve   holomorphic functions. Our explicit construction of these counterterms gives a microscopic realization of the K\"ahler ambiguities implied by the proof \GomisWY\ that the exact two-sphere partition function  \refs{\BeniniUI,\DoroudXW,\DoroudPKA} computes  the K\"ahler potential $K$ on the conformal manifold ${\cal S}_{\hbox{c}}\times {\cal S}_{\hbox{tc}}$ \refs{\JockersDK,\GomisWY}
 \eqn\kahlerAB{
 Z_{S^2}^A=e^{-K_{\hbox{tc}}}\qquad\qquad  Z_{S^2}^B=e^{-K_{\hbox{c}}}\,.}
 For some followup work see e.g. \refs{\ParkNN,\BonelliMMA}.

One well-known way in which the partition function can turn into a section is if there are 't Hooft anomalies. Then, the partition function transforms nontrivially under gauge transformations. Ordinary derivatives of the partition function then yield gauge non-invariant quantities (which can be fixed by  
the Bardeen-Zumino procedure~\BardeenPM, or equivalently, anomaly inflow~\CallanSA). Here we encounter a reminiscent situation. The partition function becomes a nontrivial section. As a result, various derivatives of the partition function need to be supplemented with an appropriate connection that for holomorphic derivatives transforms as $\CA\rightarrow\CA+\del\CF(\lambda)$ under K\"ahler transformations. It would be nice to understand if the analogy with ordinary anomalies is deeper than these superficial similarities. 
 
 In section~3 we also give a new elementary proof of \kahlerAB\ that follows from   supersymmetry Ward identities and therefore does not require the existence of a Lagrangian description or  localization of the partition function.  We also show  that the two-sphere partition function of ${\cal N}=(1,1)$ SCFTs  is not  a universal, unambiguous observable (other than capturing the A-type anomaly). We construct  a local ${\cal N}=(1,1)$ supergravity counterterm that    changes the finite part of the partition function arbitrarily, similar to what we found for even-dimensional nonsupersymmetric CFTs.

In section~4 we discuss ${\cal N}=1$ SCFTs on $S^4$.  The exactly marginal parameters again give rise to a K\"ahler manifold~\AsninXX.  We argue that in this case the partition function on the four-sphere does not have a preferred, unambiguous  nontrivial continuum limit. We show this by constructing a finite  ${\cal N}=1$ superymmetric supergravity  counterterm that includes an arbitrary real function $F$ of the moduli $\lambda^i$, which now are the bottom components of chiral multiplets $\Phi^i$
 \eqn\countertermintro{
\int d^2\Theta\, \varepsilon(\bar{\cal D}^2-8R) R\bar R\,  F(\Phi^i,\bar\Phi^{\bar i} )\,.} 
Evaluated on $S^4$, this shifts the finite part of the partition function by an arbitrary function of the moduli.

In section~5 we discuss ${\cal N}=2$ SCFTs on $S^4$. In this case we argue in two different ways (one of them being supersymmetric localization and the other by using an explicit supersymmetric regularization of the second term in~\expandZ) that 
\eqn\maini{ 
Z_{S^4}=e^{K/12}\,,}
where $K$ is the K\"ahler potential on the space of exactly marginal deformations $\CS$.  That is, the  four-sphere partition function   becomes physical if the counterterms are restricted to be ${\cal N}=2$ locally supersymmetric, and is only subjected to K\"ahler ambiguities. The partition function is again  a section rather than a function, where the transition functions are holomorphic plus anti-holomorphic in the moduli. Note the sign difference of the exponent~\maini\ with respect to~\kahlerAB.

One can therefore, in principle,   use the localization computation on $S^4$ ~\PestunRZ\  to compute the  exact Zamolodchikov metric on ${\cal N}=2$ conformal manifolds, extending what has been done for two-dimensional $\CN=(2,2)$ SCFTs.  It would be interesting to see if one can also derive~\maini\ by generalizing the argument 
of~\GomisWY\ (i.e. geometrically deforming the sphere and using the $tt^*$ equations~\CecottiME) to four dimensions (the generalization of the $tt^*$ equations should be along the lines of~\PapadodimasEU).  One may also try to  extend to four dimensions our supersymmetry Ward identity proof in section~3. It would also be very nice to explicitly construct  the four-dimensional $\CN=2$ supergravity analog of the holomorphic counterterm~\localtwod, perhaps using the tools of~\refs{\deWitZA,\ButterLTA}. Finally, it would be interesting to understand whether the conformal manifold of four-dimensional $\CN=2$ SCFTs has extra geometric structure beyond K\"ahler (additional geometric structure is known to exist in $\CN=(2,2)$ two-dimensional SCFTs with $c=9$, see e.g.~\refs{\SeibergPF,\CecottiQN,\StromingerPD}) and also~\refs{\BizetUUA} for $c=12$.
 
 In this paper we have confined ourselves to discussing conformal field theories and their sphere partition functions. However, some of our results can be relevant for the interpretation of  partition functions of gapped theories as well. For example, if we study a massive theory with a gap at the scale $M$ on a manifold with typical scale $R$ much larger than the the scale associated to the gap $R\gg 1/M$, then the partition function can be organized as a series expansion in $(RM)^{-1}$ with the various terms in the expansion corresponding to local terms in the action for the background fields. Therefore, supergravity counterterms of the type discussed in this paper can be useful to understand the leading terms in the expansion of partition functions of massive supersymmetric theories.

\newsec{${\cal N}=(2,2)$ Theories and Supersymmetric Two-Spheres}

Superconformal ${\cal N}=(2,2)$ theories in flat two-dimensional space posses a $U(1)_V\times U(1)_A$ R-symmetry group. Exactly marginal operators are superconformal descendants of operators in the chiral and twisted chiral rings which  carry charge $(2,0)$ and $(0,2)$ under $U(1)_V\times U(1)_A$.
 We can view the coupling  constants for these operators as charge $(0,0)$ background chiral superfields and charge $(0,0)$ background twisted chiral superfields, respectively. 
The conformal manifold of ${\cal N}=(2,2)$ superconformal field theories is K\"ahler and the Zamolodchikov metric takes a factorized form $\CS_{\hbox{c}}\times \CS_{\hbox{tc}}$: locally it is the sum of the 
metric in the chiral and twisted chiral directions.\foot{One can derive this as follows. Regardless of supersymmetry, in any two-dimensional CFT, promoting the exactly marginal parameters $\lambda^i$ to background fields, one has an admissible conformal anomaly~\OsbornGM \eqn\anomalytwod{\delta_\sigma W[g_{mn},\lambda^i]=\int d^2x\sqrt g\, \sigma\left(g^{mn}C_{kl}(\lambda)\del_m \lambda^k\del_n \lambda^l\right)~,}
where we omitted the usual A-type anomaly, as we have already discussed it in the introduction. $C_{kl}$ is some symmetric tensor that depends on the exactly marginal parameters. $C_{kl}$ is proportional to the Zamolodchikov metric on the space of CFTs. This is because the momentum space two-point function of exactly marginal operators looks like $p^2\log(p^2)$, which clearly contains a rescaling anomaly. Such a ``nonlinear sigma model''~\anomalytwod\ is supersymmetrized as usual, by replacing the two-derivative term with a K\"ahler function for the background superfields.  
Hence, the total space of exactly marginal couplings is K\"ahler. Since the top component of the product of a chiral field with twisted chiral fields is a total derivative, this leads to a factorized Zamolodchikov metric.
 }

Such conformal field theories can be placed on $S^2$ by a stereographic transformation. The spinors that generate the ${\cal N}=(2,2)$ superconformal transformations are conformal Killing spinors $\epsilon$ and $\tilde\epsilon$, which satisfy 
\eqn\conformaltwod{\eqalign{\nabla_m\epsilon=\gamma_m\eta~,\qquad \nabla_m\tilde\epsilon=\gamma_m\tilde\eta~.}}
Here, $\gamma_m$ are the curved space Dirac matrices, and we  define the projection operators $P_{\pm}=\half\left(1\pm\gamma_3\right)$ onto the Weyl spinors, e.g. $P_\pm\zeta=\zeta_{\pm}$. The charges of the spinors under the $R$-symmetry group are as follows
\eqn\charges{\matrix{  &  U(1)_V & U(1)_A  \cr   \epsilon_+,\eta_- & 1 & 1  \cr \epsilon_-,\eta_+ & 1 & -1\cr 
\tilde\epsilon_-,\tilde\eta_+ & -1 & 1\cr
\tilde\epsilon_+,\tilde\eta_- & -1 & -1
 }}
The equations~\conformaltwod\ have an eight-complex-dimensional space of solutions on $S^2$. These furnish four doublets of the $SU(2)$  isometry group of the two-sphere.

The superconformal algebra on  $S^2$ can be linearly realized on supermultiplets. For our purposes of analyzing the conformal manifold of ${\cal N}=(2,2)$ superconformal field theories, it will suffice to consider a chiral multiplet of charge $(-r_V,0)$ and a twisted chiral multiplet of charge $(0,-r_A)$.
 The superconformal transformations of a chiral multiplet are \DoroudXW 
\eqn\chiralmult{\eqalign{ 
{\delta} \Phi&= \tilde \epsilon \psi~,
\cr
  {\delta} \psi&=
    i\del_m\Phi \gamma^m\epsilon+\tilde\epsilon F-ir_V\eta\Phi~,
\cr
   {\delta} F&=
    -i\nabla_m\psi\gamma^m\epsilon+ir_V \psi\eta~.
 }}
A superconformal invariant can be constructed from the top component of a  chiral multiplet of Weyl weight 1  ($r_V=-2$)
\eqn\invchiral{
I=\int d^2x\sqrt{g}\, F\,,}
since in this case ${\delta}  {F}=   -i\nabla_m\left( {\psi}\gamma^m {\zeta} \right)$, and the variation  integrates to zero.

The twisted chiral multiplet transformations are given by \GomisWY
\eqn\twistedconf{\eqalign{
\delta Y &= \left(\tilde{\epsilon}P_{-}-\epsilon P_{+}\right)\zeta~,
\cr
  \delta \zeta_{+} &= - P_{+} \left( i \del_m Y\gamma^m - G \right) \tilde\epsilon  +ir_A Y\tilde \eta_+~,
\cr
   \delta \zeta_{-} &= P_{-} ( i \del_mY\gamma^m - G ) \epsilon -ir_A Y \eta_-~,
\cr
  \delta G &=  i\epsilon P_{-} \slashchar{\nabla} \zeta 
	  - i \tilde\epsilon P_{+}  \slashchar{\nabla} \zeta    +i r_A \zeta P_-\eta  -ir_A  \zeta P_+\tilde \eta~.
}}
For a twisted chiral multiplet  of Weyl weight one ($r_A=-2$) the integral of the top component
\eqn\invtwistedchiral{
I=\int d^2x\sqrt{g}\, G\,}
is a superconformal invariant, since $\delta G=i \nabla_m\left(\epsilon P_-\gamma^m \zeta \right)-i \nabla_m\left(\tilde\epsilon P_+\gamma^m \zeta\right)$.

The superconformal invariants \invchiral\invtwistedchiral\ represent the exactly marginal operators that are descendants of   operators in the chiral and twisted chiral rings. The anti-chiral and anti-twisted-chiral transformations and invariants are constructed similarly. 

The superconformal superalgebra just described admits interesting ``massive subalgebras'' with four supercharges. By this we mean that massive $\CN=(2,2)$ quantum field theories can be placed on $S^2$ while respecting these subalgebras~\refs{\DoroudXW,\BeniniUI,\GomisWY,\DoroudPKA}. 

A regularization scheme can be thought of as a deformation of the theory by some massive sector. Hence, in order to understand the possible ways of regularizing the partition function $Z_{S^2}$ while preserving some supersymmetry, we need to study these massive subalgebras in detail. 
We will mainly focus on the subalgebras that preserve four supercharges, but will also pay some attention to one subalgebra that preserves two supercharges. 

 There are two inequivalent massive ${\cal N}=(2,2)$ supersymmetry algebras on the two-sphere. They are generated by supercharges in the superconformal algebra that close into the $SU(2)$ isometry of the two-sphere and a $U(1)$ subgroup of the $U(1)_V\times U(1)_A$ superconformal R-symmetry. The  algebras that preserve $U(1)_V$ and $U(1)_A$ have been denoted by $SU(2|1)_A$ and $SU(2|1)_B$   in \refs{\DoroudXW,\DoroudPKA}. They are mapped to each other by the mirror outer automorphism~\LercheUY\ of the ${\cal N}=(2,2)$ superconformal algebra. 

One can also study massive subalgebras preserving only two supercharges. We do not carry out an exhaustive analysis of this case, but only discuss one example: the massive $\CN=(1,1)$ subalgebra $OSp(1|2)$,  consisting of  an $SU(2)$ doublet of supercharges closing into the $SU(2)$ isometries of $S^2$, thus   breaking  the $R$-symmetry group completely.

 $\bullet$ $SU(2|1)_A$.  This supersymmetry algebra is obtained by restricting the conformal Killing spinors on the two-sphere \conformaltwod\ to obey
\eqn\globalsphere{
\nabla_m \epsilon={i\over 2r} \gamma_m \epsilon\qquad \nabla_m \tilde \epsilon={i\over 2r} \gamma_m \tilde\epsilon\,,}
so that
\eqn\speconf{
\eta={i\over 2r} \epsilon~,\qquad \tilde \eta={i\over 2r} \tilde\epsilon\,.}
In stereographic coordinates, where $ds^2={1\over \left(1+ {x^2\over 4r^2}\right)^2} dx^m dx^n \delta_{nm}$, the  complex  four-dimensional space of solutions is given by
\eqn\killspheretwo{\eqalign{
\epsilon={1\over \sqrt{1+{x^2\over 4r^2}}} \left(1+ {i \over 2r} x^m\Gamma_m\right) \epsilon_0~,\cr
\tilde\epsilon={1\over \sqrt{1+{x^2\over 4r^2}}} \left(1+ {i \over 2r} x^m\Gamma_m\right) \tilde\epsilon_0\,,}}
where $\Gamma_m$ denotes the flat space gamma matrices (the usual Pauli matrices), i.e. $\Gamma_m=\left({1+{x^2\over 4r^2}}\right)\gamma_m$.
The $SU(2|1)_A$ supersymmetry transformations are  the restriction of  \chiralmult\ and \twistedconf\  to these conformal Killing spinors.

  $\bullet$ $SU(2|1)_B$.  This supersymmetry algebra is obtained by restricting the conformal Killing spinors on the two-sphere \conformaltwod\ to obey
\eqn\globalsphereb{
\nabla_m \epsilon={i\over 2r} \gamma_m \tilde\epsilon~,\qquad \nabla_m \tilde \epsilon={i\over 2r} \gamma_m  \epsilon\,,}
so that
 $\varepsilon=\epsilon_++\tilde\epsilon_-$ and $\bar\varepsilon=\tilde\epsilon_++\epsilon_-$ satisfy
\eqn\spinors{
\nabla_m \varepsilon={i\over 2r} \gamma_m \varepsilon~,\qquad \nabla_m \bar \varepsilon={i\over 2r} \gamma_m \bar \varepsilon\,,
}
and
\eqn\speconfb{
\eta={i\over 2r} \tilde\epsilon~,\qquad \tilde \eta={i\over 2r}  \epsilon\,.}
The $SU(2|1)_B$ supersymmetry transformations are  the restriction of  \chiralmult\ and \twistedconf\  to these conformal Killing spinors.

$\bullet$ $OSp(1|2)$: Massive $\CN=(1,1)$. The superconformal $\CN=(1,1)$ superalgebra is embedded in the superconformal $\CN=(2,2)$ superalgebra by the restriction $\epsilon=\tilde\epsilon$ (and thus $\eta=\tilde \eta$). This superconformal algebra has four supercharges. It further has a massive $OSp(1|2)$ subalgebra that is obtained by setting $\eta={i\over 2r}\epsilon$. This has two supercharges. The chiral and twisted chiral multiplets become reducible, each breaking up into two $\CN=(1,1)$ real scalar multiplets. 

\subsec{The $S^2$ Partition Function}

Let us now consider the $S^2$ partition function of $\CN=(2,2)$ superconformal theories. Since the formal expansion~\definition\ contains divergences, we need to decide which symmetries our regulator preserves. Below we will analyze the consequences of the assumption that the symmetry preserved by the regulator is one of the massive subalgebras of the ${\cal N}=(2,2)$ superconformal algebra discussed above.

Let us begin by assuming that the physics at coincident points is $SU(2|1)_A$ invariant. That means that we are allowed to use $SU(2|1)_A$ Ward identities, even at coincident points. 
With this assumption, the first observation we make is that the chiral multiplet invariant~\invchiral\ is $SU(2|1)_A$-exact. By taking $\epsilon=0$ in \chiralmult\ we have that
\eqn\exactc{
F={1\over \|\tilde \epsilon\,\|^2}\delta \left(\tilde\epsilon^\dagger \psi\right)\,,
}
where $\|\tilde \epsilon\,\|^2\equiv \tilde\epsilon^\dagger \tilde\epsilon$, and $\tilde\epsilon$ is a  nowhere vanishing Dirac conformal Killing spinor. Therefore, the $S^2$ partition function is independent of the chiral couplings constants. This already shows that the two-sphere partition function is not completely ambiguous. The independence of the chiral couplings implies that there is no local $SU(2|1)_A$ supergravity counterterm that could reduce on $S^2$ to a function of the bottom components of the chiral couplings.

The   twisted chiral multiplet invariant \invtwistedchiral\ is more interesting. It is not exact with respect to the  $SU(2|1)_A$ subalgebra of the superconformal algebra. Indeed, in $SU(2|1)_A$ it is inconsistent to set $\tilde\epsilon_-=0$ while keeping $\tilde\epsilon_+\neq 0$, as the two chiralities are linked by the $SU(2|1)_A$ Killing spinor equation \globalsphere. The $SU(2|1)_A$-invariant partition function may thus depend on the twisted chiral moduli.

Even though the top component of the  twisted chiral multiplet $G$ is not exact, its integral is {\it almost} so.     From \twistedconf\ we see that  
\eqn\exactal{
\delta\zeta_+=  -i  \slashchar{\nabla}\left(Y\,\tilde\epsilon_-   \right)+ G\, \tilde\epsilon_+~.}
If $\tilde \epsilon_+$ had been nowhere vanishing, then the integrated top component would have been exact. But $\tilde \epsilon_+$ does vanish. Without loss of generality, we can take the zero to be at $x=0$ (the North Pole). This corresponds to choosing $\tilde\epsilon_{0+}=0$  in \killspheretwo\ so that
\eqn\killvan{\eqalign{
\tilde\epsilon_+&={i/2r\over \sqrt{1+{x^2\over 4r^2}}}    x^m\Gamma_m   \tilde\epsilon_{0-}~,\cr
\tilde\epsilon_-&={1\over \sqrt{1+{x^2\over 4r^2}}}\tilde\epsilon_{0-}~.
}}
We note that the insertion of the bottom component of a twisted chiral multiplet $Y$ is invariant under this supersymmetry transformation if the
operator is inserted at the North Pole of the two-sphere (and if $\epsilon_-$ also vanishes at the North Pole).

We can focus on the physics near the point where the spinor vanishes as follows. Introducing complex coordinates $z=x^1+ix^2$ and $\bar z=x^1-ix^2$ (and using that $\nabla_z \tilde\epsilon_+=0$ and $\nabla_z\left( \tilde\epsilon_+^{\,\dagger}\over  ||\tilde \epsilon_+||^2 \right) =0$), after some simple algebra we arrive at  
\eqn\Galm{
G=\delta\left({\tilde\epsilon^\dagger_+  \zeta_+\over ||\tilde \epsilon_+||^2}\right)+2i \nabla_z\left({1+{z \bar z\over 4r^2}\over ||\tilde \epsilon_+||^2}\tilde\epsilon_+^\dagger \Gamma_1 \tilde\epsilon_- Y \right)\,.
} 
This expression is only valid away from the North Pole.

We rewrite the supersymmetric invariant as
\eqn\susyinvpart{
\int d^2x \sqrt{g}\, G=\int_{D_S} d^2x \sqrt{g} \,G+\int_{D_N} d^2x \sqrt{g} \,G\,,}
where $D_N=\{x | x^2\leq R^2\}$ and $D_S=\{x | x^2\geq R^2\}$. 
 Using the expression \Galm, which is regular in $D_S$, we get 
\eqn\north{
\int_{D_S} d^2x \sqrt{g}\, G= \int_{D_S} d^2x \sqrt{g} \delta\left({\tilde\epsilon^{\,\dagger}_+  \zeta_+\over ||\tilde\epsilon_+||^2}\right)+2i  \int_{D_S} d^2x \sqrt{g} \, \nabla_z\left({1+{z \bar z\over 4r^2}\over ||\tilde \epsilon_+||^2}\tilde\epsilon_+^\dagger \Gamma_1 \tilde\epsilon_- Y \right)}
Inside a correlation function with other $\delta$-closed operator insertions, the first term vanishes  and the entire contribution from $D_S$ comes from the second, total derivative term. 
 Ultimately we will take the limit that $R\rightarrow 0$, so that the contribution from  $D_N$ is vanishingly small. We thus need to evaluate the total derivative term in~\north. Using polar coordinates, Stokes' theorem, and the identities  
\eqn\formulabil{
||\tilde\epsilon_+||^2={1\over 4r^2}{ z \bar z\over  1+{z \bar z\over 4r^2}} ||\tilde \epsilon_{0-}||^2 
\qquad  
\tilde\epsilon_+^\dagger \Gamma_1 \tilde\epsilon_- =-{i\over 2r}{ z\over  1+{z \bar z\over 4r^2} }||\tilde \epsilon_{0-}||^2~,
 }
 yields	
\eqn\calcul{
\int d^2x \sqrt{g}\, G =-  2r \, \hbox{lim}_{R\rightarrow 0}  \int_0^{2\pi}d\theta {Y(R,\theta)\over 1+{R^2 \over 4r^2}}=-4\pi rY(0)=-4\pi r Y(N)\,.
}
An analogous analysis for the top component of the anti-twisted chiral multiplet yields 
\eqn\calculanti{
\int d^2x \sqrt{g}\, \bar G =4\pi r\bar Y(\infty)=4\pi r \bar Y(S)\,.
}
Therefore, the integrated twisted chiral (anti-twisted chiral) multiplet top component  inside a correlation function inserts the bottom component of the twisted chiral (anti-twisted chiral) multiplet at the North (South) Pole of the two-sphere.
We will now use these formulae to prove that $Z_{S^2}^A=\exp(-{K}_{\hbox{tc}})$.
 
 Differentiating  the partition function twice with respect to twisted chiral moduli we get
 \eqn\twisttwopt{
 \partial_i\partial_{\bar j} \log Z_{S^2}^A={1\over \pi^2}\vev{\int_{S^2}d^2x \sqrt{g}\,  G_i(x)\int_{S^2}d^2y\sqrt{g} \, \bar {G}_{\bar j}(y)}\,.}
 Using \calcul\calculanti\ we arrive at  
 \eqn\zamolo{
  \partial_i\partial_{\bar j} \log Z_{S^2}^A=-(4r)^2\vev{Y_i(N)\bar Y_{\bar j}(S)}=-(2r)^4\vev{G_i(N)\bar G_{\bar j} (S)}=-\partial_i\partial_{\bar j}{K}_{\hbox{tc}}\,.
  }
In completing the proof we have used  a supersymmetry Ward identity relating the two-point functions of the top and the bottom components of a chiral multiplet, i.e. $\vev{G_i(N) \bar G_{\bar j}(S)}={1\over r^2} \vev{Y_i(N)\bar Y_{\bar j}(S)}$, and the fact that the metric on the conformal manifold of ${\cal N}=(2,2)$ superconformal field theories is  
K\"ahler, i.e. $g_{i\bar j}=\partial_i\partial_{\bar j}{K}$. Therefore, up to a holomorphic ambiguity which will be discussed in detail in the next subsection,
\eqn\zamoloi{Z_{S^2}^A=e^{-K_{\hbox{tc}}}~.}
 
If we assume that the physics at coincident points is $SU(2|1)_B$ invariant, the analysis is similar to what we have done above, essentially only exchanging the role of chiral and twisted chiral multiplets.  The twisted chiral invariant \invtwistedchiral\ is $SU(2|1)_B$-exact. The chiral invariant \invchiral\ is not $SU(2|1)_B$-exact, again, because one cannot construct it as a variation of a fermion without encountering some zero of a Killing spinor.
Repeating the steps above we arrive at
\eqn\kahlerB{
  \partial_i\partial_{\bar j} \ln Z_{S^2}^B= -(4r)^2\vev{\phi_i(N)\bar \phi_{\bar j}(S)}=-(2r)^4\vev{F_i(N)\bar F_{\bar j} (S)}= -\partial_i\partial_{\bar j}K_{\hbox{c}}\,.}
Thus, up to a holomorphic ambiguity which we will discuss below, 
\eqn\zamoloii{Z_{S^2}^B=e^{-K_{\hbox{c}}}~.}

Let us now discuss the consequences of preserving the massive $\CN=(1,1)$  superalgebra $OSp(1|2)$  described in the previous subsection. The exactly marginal operators are given by the top components of  $\CN=(1,1)$ real scalar multiplets with Weyl weight $1$. This  multiplet is 
just the real version of the $\CN=(2,2)$ chiral multiplet \chiralmult, and therefore the $\CN=(1,1)$ superconformal invariant is \invchiral, but with  $F$ now being real. Let us now consider the sphere partition function regulated in an $OSp(1|2)$  invariant way. One can attempt to repeat the analysis performed for $SU(2|1)_A$ and $SU(2|1)_B$. However, we find that the integrated  two-point function of the  top component of a real scalar multiplet does not reduce to the unintegrated two-point function of the bottom components, in contrast to the ${\CN=(2,2)}$ analysis. While it is still true that the top component is not supersymmetry exact (it is  of the form in \exactal), now the bottom component of the ${\CN=(1,1)}$ multiplet is not $\delta$-closed,  since   Dirac Killing spinors on $S^2$ are nowhere vanishing.   Note that in our $\CN=(2,2)$ derivation of \zamolo\ we used in a crucial way that the insertion of the bottom component of the $\CN=(2,2)$ multiplets at the poles  is $\delta$-closed under a supersymmetry transformation.  
In the next subsection we will construct an  explicit ${\cal N}=(1,1)$ supergravity counterterm involving an arbitrary function of the moduli, which also sit in bottom components of ${\cal N}=(1,1)$ real chiral multiplets. We conclude that the partition function of ${\cal N}=(1,1)$  SCFTs on $S^2$ is scheme dependent and thus ambiguous.  We will reach a very similar conclusion for the sphere  partition function of four-dimensional $\CN=1$ SCFTs in section~4.

Before we turn to the holomorphic ambiguity afflicting the partition functions $Z_{S^2}^{A,B}$, let us make some comments. We discussed what happens if the theory is regulated in a manner that preserves $SU(2|1)_{A,B}$. But what if we could find a regulator that preserves a bigger symmetry group, i.e. the full superconformal group which has eight supercharges on $S^2$? 
Clearly, such a regulator cannot exist. The reason is simply that if it had existed, using two different subgroups of the superconformal symmetry group, we would have arrived at two different results~\zamoloi,\zamoloii. In fact, if all the eight supercharges are available to us, then one can formally prove that the chiral invariant~\invchiral\ and the twisted chiral invariant~\invtwistedchiral\ are both superconformal exact. Hence, one would be led to the conclusion that the partition function is independent of the exactly marginal parameters. The fact that one can get contradictory answers simply means that the full $\CN=(2,2)$ superconformal group cannot be preserved. This is reminiscent of an anomaly.

\subsec{K\"ahler   Ambiguity Counterterm and $\CN=(1,1)$ Supergravity Counterterm}  

The K\"ahler potential admits  the following well-known ambiguity
\eqn\ambigui{
K(\lambda,\bar \lambda)\rightarrow K(\lambda,\bar \lambda)+ \CF(\lambda)+\bar \CF(\bar \lambda)\,,}
which leaves the metric $g_{i\bar j}=\partial_i\partial_{\bar j}{K}$ invariant.
The fact that the two-sphere partition function computes the K\"ahler potential implies that there should
be a finite, local  ${{\cal N}=(2,2)}$ supergravity counterterm that when evaluated on the two-sphere captures the
K\"ahler ambiguity \ambigui. For example, if we describe the same conformal field theory in two different duality frames, the partition functions may differ by some holomorphic (plus anti-holomorphic) function of the moduli. This means that the partition function is a section on the conformal manifold.

We proceed to construct this local supergravity counterterm explicitly.

There are two minimal versions of $\CN=(2,2)$ supergravity.  They differ by the nature of the $U(1)$ $R$-symmetry that is gauged (vector or axial), and are mapped into each other under the $Z_2$ mirror automorphism.   The rigid limits of these two supergravities correspond to preserving $SU(2|1)_A$ or $SU(2|1)_B$ on the two-sphere.

We begin with the $SU(2|1)_B$ theory. 
The ambiguity \ambigui\ depends on the couplings to the exactly marginal operators from the chiral ring. They sit in background chiral multiplets of vanishing  $U(1)_A$ $R$-charge. The supergravity counterterm supersymmetrizes the product of the Ricci scalar curvature $\CR$ with the  sum of an arbitrary  holomorphic function and an anti-holomorphic function of  chiral multiplets
\eqn\bosonic{
\int d^2x \sqrt{g}\, \CR \left(\CF(\lambda^i)+\bar\CF(\bar\lambda^{\bar i})\right)\,.
}
 The relevant local counter term is obtained from the $SU(2|1)_B$  supergravity action\foot{We thank N.~Seiberg for a discussion.}~\KetovES\
\eqn\sugra{
\int d^2 x d^2\Theta\,\varepsilon^{-1}R\,\CF(\Phi^i)+c.c~,
}
where $\CF(\Phi^i)$ is a general holomorphic function of the chiral multiplets, $\varepsilon^{-1}$ is the chiral density superspace measure and $R$ is a chiral superfield whose  bottom component  is a complex scalar auxiliary field that belongs to the $SU(2|1)_B$ supergravity multiplet. The spacetime scalar curvature $\CR$ sits in the top component of the superfield $R$.
When~\sugra\   is evaluated in the $SU(2|1)_B$ supersymmetric  two-sphere background and the superfields $\Phi^i$ are replaced by the constant   exactly marginal parameters $\lambda^i$, we reproduce \bosonic. This establishes the K\"ahler ambiguity for the $SU(2|1)_B$ theory.

Similarly, for the $SU(2|1)_A$ theory one has to consider \eqn\sugrai{{\int d^2xd\Theta^+d\t\Theta^-\h\varepsilon^{\,-1}F \CF(\Omega^i)+c.c~,}} where $ \CF(\Omega^i)$ is an arbitrary holomorphic function of the twisted multiplets, and $\h\varepsilon^{-1}$ is the twisted chiral density superspace measure. $F$ is a twisted chiral superfield that contains as its lowest component a complex scalar auxiliary field that belongs to the  $SU(2|1)_A$-supergravity multiplet and the scalar curvature $\CR$ in its top component. The rigid limit of this coupling evaluated on supersymmetric backgrounds was recently considered in \ClossetPDA. This supergravity counterterm evaluated  in the $SU(2|1)_A$ supersymmetric  two-sphere background yields the marginal counterterm \bosonic,  where $\lambda^i$ now stand for twisted chiral exactly marginal couplings. 

Hence, the finite piece of the two-sphere partition function has a holomorphic plus an anti-holomorphic ambiguity. The pieces that cannot be shifted away by a holomorphic and an anti-holomorphic function of the exactly marginal parameters are physical and calculable, as long as our regularization scheme respects $SU(2|1)_A$  or $SU(2|1)_B$, as explained in the previous subsection. 

 We now construct a two-dimensional $\CN=(1,1)$ supergravity counterterm that depends on an arbitrary function of the moduli. This is an $\CN=(1,1)$  
  local supersymmetrization of 
 \eqn\localone{\int d^2x\sqrt g\, \CR\, F(\lambda^i)\,.}
 We write down this counterterm by coupling    $\CN=(1,1)$   supergravity   to real scalar multiplets $\Phi^i$.  The moduli $\lambda^i$ are the bottom components of $\Phi^i$ while the supergravity multiplet contains the graviton, the gravitino and a real auxiliary field $B$. The superspace description of this $\CN=(1,1)$ supergravity counterterm is  
\eqn\super{\int d^2xd^2\Theta E^{-1}R\,F(\Phi^i)\,,}
 where $E^{-1}$ is  the density measure superfield and $R$ is a  real  superfield, which has a complex scalar  $R|=B$ as its bottom component and contains the scalar curvature ${\cal R}$ in its top component \GatesHY. This supergravity counterterm evaluated  in the $OSp(1|2)$  supersymmetric  two-sphere background yields the marginal counterterm \localone. Therefore, the sphere partition function of $\CN=(1,1)$ SCFTs  is not much different from the sphere partition function of nonsupersymmetric CFTs, for which     the $S^2$ partition function does not have a preferred nontrivial continuum limit. As we shall see in section~4, we will reach the same conclusion when we study the sphere partition function of four-dimensional $\CN=1$ SCFTs.

\subsec{An Explicit Regularization}

Let us for a moment forget about supersymmetry and examine a little more closely the quadratic term in the expansion~\definition\ for the partition function, i.e.~$\CO=1$. This is proportional to the double integral 
\eqn\inte{
\vev{\int_{S^d}d^dx \sqrt{g}\,   O_i(x)\int_{S^d}d^dy\sqrt{g} \, O_{j}(y)}\,.}
The two-point function on $S^d$ can be obtained from the one in flat space by a Weyl transformation: 
\eqn\twopoint{
\vev{O_i(x) O_{j}(y)}_{S^d}={g_{ij}\over
d(x,y)^{2\Delta}}\,,}
where 
\eqn\dist{
d(x,y)={|x-y|\over \sqrt{1+{x^2\over 4r^2}}\sqrt{1+{y^2\over 4r^2}}}\,,}
and since the operators are exactly marginal $\Delta=d$. The integral~\inte\ is ultraviolet divergent. It can be regulated
by setting $\Delta=d-\epsilon$. For sufficiently large and positive $\epsilon$ the integral converges and evaluates to
\eqn\integra{\eqalign{
\vev{\int_{S^d}d^dx \sqrt{g}\,  O_i(x)\int_{S^d}d^dy\sqrt{g} \,  O_{j}(y)}&=g_{ij} {\hbox{vol}(S^d)\hbox{vol}(S^{d-1})\over 2^{d+1}} {\Gamma(d/2)\Gamma(-d/2+\epsilon)\over \Gamma(\epsilon)}\,.}}

Setting $d=2$,  in the $\epsilon\rightarrow 0$ limit we get 
that \eqn\twodim{
\vev{\int_{S^2}d^2x \sqrt{g}\,  O_i(x)\int_{S^2}d^2y\sqrt{g} \,  O_{j}(y)}=-\pi^2 g_{ij}~. }
In the case without any supersymmetry, this procedure of regularizing~\inte\ by continuing the dimension of the operator is just one of many possible regularization schemes and since the answer is not renormalization scheme invariant, there is no particularly special meaning to~\twodim. 

Now let us consider this regularization scheme in the context of $\CN=(2,2)$ SCFTs. We see that~\twodim\ gives an incorrect answer in some situations! Indeed, we have shown above that, if we preserve $SU(2|1)_A$,  the partition function is independent of the chiral couplings. However~\twodim\ does not distinguish the chiral from the twisted chiral couplings. 

The resolution is, of course, that the procedure of analytically continuing the dimension of chiral operators is inconsistent with $SU(2|1)_A$. This is because if one changes the dimension of a chiral operator in the superpotential, one necessarily breaks $U(1)_V$ and therefore $SU(2|1)_A$. However, this procedure is perfectly $SU(2|1)_A$ invariant  for the twisted chiral operators, and since we have already explained that the answer is unique,~\twodim\ must give the right result in this case.
Indeed,~\twodim\  is completely consistent with the formula derived in \zamolo, which implies \zamoloi. The mirror symmetric statement is that this regularization scheme preserves $SU(2|1)_B$ for chiral operators   while it breaks it for twisted chiral operators, 
 and therefore leads to  \zamoloii. We will use this derivation via analytic continuation in the conformal dimension of the exactly marginal operators  again in section~5, when we discuss four-dimensional $\CN=2$  SCFTs.

\newsec{${\cal N}=1$ Field Theories on $S^4$}

 We now wish to study the dependence of the four-sphere partition function of ${\cal N}=1$ superconformal field theories on their exactly-marginal couplings. The exactly-marginal couplings parameterize a K\"ahler manifold~\AsninXX, and they are commonplace in ${\cal N}=1$ superconformal theories~\LeighEP. These theories have a $U(1)_R$ R-symmetry. The exactly marginal operators of ${\cal N}=1$ superconformal field theories   are descendants of operators in the chiral ring; they are the top component of a chiral multiplet with R-charge $2$ (and Weyl weight 3).   We note that not all such operators are exactly marginal in four-dimensional ${\cal N}=1$ superconformal field theories. This is unlike in 
unitary two-dimensional ${\cal N}=(2,2)$ superconformal field theories with a normalizable vacuum state, where superconformal descendant operators of the type described in the previous section are necessarily exactly marginal.\foot{The origin of this difference is as follows: as explained in~\GreenDA, a marginal operator can cease to be exactly marginal if there is a dimension $d-2$ real multiplet that can ``eat'' our marginal operator and thereby become a longer multiplet. In four-dimensional $\CN=1$ theories this is possible using conserved current multiplets. Since in two dimensions $d-2=0$, in unitary theories there is no candidate multiplet to ``eat'' our marginal operator. Therefore, it has to be exactly marginal.}
  
 ${\cal N}=1$ superconformal field theories can be canonically placed on the round four-sphere. The superconformal transformations are generated by a Dirac conformal Killing spinor  on $S^4$, which obeys
 \eqn\cksfour{
 \nabla_m\epsilon=\gamma_m\eta\,.}
 Under superconformal transformations  the components of a chiral multiplet of Weyl weight $w$ (and R-charge ${2\over 3}w$) transform as~\FreedmanZZ\
  \eqn\superchiralfour{\eqalign{
  \delta Z &={1\over \sqrt{2}}\bar\epsilon\chi_L~,\cr
  \delta \chi_L &={1\over \sqrt{2}}\left(\slashchar{\nabla}Z \epsilon_R+F \epsilon_L\right)+\sqrt{2}w Z \eta_L~,\cr
  \delta F &={1\over \sqrt{2}}\bar\epsilon\slashchar{\nabla}\chi_L+\sqrt{2}(1-w)\bar\eta\chi_L={1\over \sqrt{2}}\nabla_m\left(\bar\epsilon\gamma^m\chi_L\right)+\sqrt{2}(3-w)\bar\eta\chi_L\,,}}
     where $\bar\epsilon \equiv \epsilon^T C$, and   $C$ is the charge conjugation matrix. We have also defined chiral spinors with the projectors $P_L={1\over 2}(1+\gamma_*)$ and $P_R={1\over 2}(1-\gamma_*)$, so that $P_{L,R}\epsilon=\epsilon_{L,R}$.  
     
We can construct the following ${\cal N}=1$ superconformal invariant using a chiral multiplet of  Weyl weight $w=3$ 
       \eqn\invarfourd{
 I=\int d^4x \sqrt{g} F~.}
This represents a general marginal operator in the superconformal field theory. From now on, we will only discuss the exactly marginal ones. The parameters multiplying these operators, the coordinates in the conformal manifold, are realized as the lowest component of background chiral multiplets with vanishing $R$-charge.
 
 The massive  $OSp(1|4)$ subalgebra of the   ${\cal N}=1$ superconformal algebra  is realized by the following conformal Killing spinors on $S^4$
\eqn\fdno{
\nabla_m \epsilon= {i\over 2r} \gamma_m \epsilon\,,}
which corresponds to choosing
\eqn\ettafour{
\eta={i\over 2r} \epsilon\,.}
The supersymmetry transformation generated by these spinors anti-commute to give the $SO(5)$ isometry of the four-sphere, hence, projecting out the conformal and R-symmetry generators in the  ${\cal N}=1$ superconformal algebra. This is why one can put massive ${\cal N}=1$ theories on $S^4$ preserving~\fdno.  The action of this subalgebra on the chiral multiplet~\superchiralfour\ is of course induced from the action of the full superconformal group, restricted to the spinors~\fdno.

One can attempt to repeat our analysis in two dimensions and aim to write the invariant~\invarfourd\ as an $OSp(1|4)$ supersymmetry variation of a fermion modulo a total derivative and a zero. However, one quickly encounters a geometrical obstruction. One finds an additional term, 
proportional to $[\xi, \xi^*]_m$, where $\xi^m=\half\bar \epsilon \gamma^m \epsilon$ is a complex Killing vector. This commutator cannot be put to zero on $S^4$ (loosely speaking, the $OSp(1|4)$ algebra contains only complexified Killing vectors not commuting with their complex conjugate).

 Therefore, the invariant \invarfourd\ is not $OSp(1|4)$-exact (even if one ignores zeroes of spinors and total derivatives).  In fact, we   now prove that the dependence of the sphere partition function of ${\cal N}=1$ superconformal field theories on moduli is scheme dependent. This is therefore analogous to the situation in non-supersymmetric theories.

There  is a finite, local ${\cal N}=1$ supergravity counterterm that depends on an arbitrary real function of the chiral  and anti-chiral multiplets. This counterterm once evaluated on the $OSp(1|4)$ invariant   four-sphere is non-vanishing and explicitly demonstrates that the dependence on the moduli of the partition function is ambiguous.
Denoting chiral fields by $\Phi^i$ and using the conventions of~\WessCP, the relevant supergravity counterterm is
\eqn\counterterm{
-{2\over3}\int d^4x\int d^2\Theta\, \varepsilon({\bar D}^2-8R)R\bar R F(\Phi^i,\bar\Phi^{\bar i})~.} 
The supergravity multiplet contains, in addition to the graviton and the gravitino, two auxiliary fields - a complex scalar $M$ and a real vector $b_\mu$. $\varepsilon$ is the chiral superspace measure and is the superfield that contains the square root of the determinant of the metric in its bottom component. The chiral superfield $R$ has the auxiliary field $M$ as its lowest component. After setting the gravitino to zero, we get:
\eqn\jkjkj{\eqalign{
&\varepsilon={1\over2}\sqrt g-{1\over2}\sqrt g\bar M\Theta^2~,
\cr
&R=-{1\over6}M-{1\over6}\Theta^2\Big(-{1\over2}{\cal R} +{2\over3}M\bar M+{1\over3}b_\mu b^\mu-i\nabla_\mu b^\mu\Big)~.
}}

Manifolds that preserve four supercharges obey the integrability conditions~\FestucciaWS:
\eqn\integrability{\eqalign{
&{3\over2}{\cal R}-b_\mu b^\mu-2M\bar M=0~,
\cr
&\nabla_\mu b^\mu=0~.
}}
Using \integrability, only the lowest component of $R$ remains.
For a chiral superfield $\Omega$ with lowest component $\omega$, we have
\eqn\vnvn{
(\bar{\cal D}^2-8R)\bar R\bar\Omega=-{2\over9}M\bar M\bar\omega+\Theta^2\Big({2\over3}\nabla_\mu\partial^\mu(\bar M\bar{\omega})+{{4i}\over9}b_\mu\partial^\mu(\bar M\bar{\omega})\Big)~,
}
where we have set the higher components of $\Omega$ to zero since they vanish on the supersymmetric sphere (in addition, we have set the gravitino to zero).

The background fields for $S^4$ are $M=\bar M=-{{3i}\over r}$ and $b_\mu=0$, which solves~\integrability. After a little algebra, we then obtain that~\counterterm\ evaluates to 
\eqn\hhgjhg{
 {1\over{r^4}}\int d^4x \sqrt g   F(\lambda^i,\bar\lambda^{\bar i})
}
on the supersymmetric sphere.
Thus,  \counterterm\ supersymmetrizes the finite counterterm \hhgjhg\ for an arbitrary function of the exactly marginal parameters. 
  This renders the sphere partition function of four-dimensional $\CN=1$ SCFTs ambiguous.

   \newsec{${\cal N}=2$ Field Theories on $S^4$}
   
   We now turn our attention to the study of the four-sphere partition function of ${\cal N}=2$ superconformal field theories.
   These theories have an $SU(2)_R\times U(1)_R$ R-symmetry.
   An exactly marginal operator  in such a  theory is  realized as a superconformal descendant of the bottom component of a  chiral multiplet of $U(1)_R$ R-charge $2$ (and Weyl weight $2$).
   
We consider now placing   ${\cal N}=2$ superconformal field theories on $S^4$. The ${\cal N}=2$ superconformal transformations are parametrized by an $SU(2)_R$ R-symmetry doublet of left chiral conformal Killing spinors $\epsilon^i$ and   right chiral conformal Killing spinors $\epsilon_i$
\eqn\ckstwo{
\nabla_m \epsilon^i=\gamma_m\eta^i\qquad \nabla_m \epsilon_i=\gamma_m\eta_i\,.}
   The supersymmetry transformations of a four dimensional ${\cal N}=2$ chiral multiplet of Weyl weight $w$ are given by~\Toine
\eqn\SUSY{\eqalign{
\delta A = & {1\over 2}\bar \epsilon ^i\Psi _i\,,  \cr
     \delta \Psi_i =&\,\slashchar{\nabla} A\epsilon_i + {1\over 2}B_{ij}\,\epsilon^j +
 {1\over 4}   \gamma^{ab} F_{ab}^- \,\varepsilon_{ij} \epsilon^j + 2w
  A\,\eta_i\cr =& \slashchar{\nabla} (A\epsilon_i )+ {1\over 2}B_{ij}\,\epsilon^j +
 {1\over 4}   \gamma^{ab} F_{ab}^- \,\varepsilon_{ij} \epsilon^j + \left({2w}-4\right)A \eta_i 
   \cr
  \delta B_{ij} =&\, \bar\epsilon_{(i} \slashchar{\nabla} \Psi_{j)} - \,
  \bar\epsilon^k \Lambda_{(i} \,\varepsilon_{j)k} + 2(1-w)\,\bar\eta_{(i}
  \Psi_{j)} \,, \cr
  \delta F_{ab}^- =&\, {1\over 4} 
  \varepsilon^{ij}\,\bar\epsilon_i\slashchar{\nabla}\gamma_{ab} \Psi_j+
   {1\over 4}  \bar\epsilon^i\gamma_{ab}\Lambda_i
  - {1\over 2} (1+w)\,\varepsilon^{ij} \bar\eta_i\gamma_{ab} \Psi_j \,,
 \cr
  \delta \Lambda_i =&\,- {1\over 4} \gamma^{ab}\slashchar{\nabla}F_{ab}^-
   \epsilon_i  -{1\over 2}\slashchar{\nabla}B_{ij}\varepsilon^{jk} \epsilon_k +
  {1\over 2}C\varepsilon_{ij}\,\epsilon^j
  -(1+w)\,B_{ij}
  \varepsilon^{jk}\,\eta_k +  {1\over 2}  (1-w)\,\gamma^{ab}\, F_{ab}^-
    \eta_i \, \cr
    =&\,- {1\over 4} \gamma^{ab}\slashchar{\nabla}(F_{ab}^-
   \epsilon_i)  -{1\over 2}\slashchar{\nabla}B_{ij}\varepsilon^{jk} \epsilon_k +
  {1\over 2}C\varepsilon_{ij}\,\epsilon^j
  -(1+w)\,B_{ij}
  \varepsilon^{jk}\,\eta_k +  {1\over 2}  (3-w)\,\gamma^{ab}\, F_{ab}^-
    \eta_i\,,\cr
    \delta  C =& -  \varepsilon^{ij} \bar\epsilon_i
\slashchar{\nabla} \Lambda_j
                            + 2 w\varepsilon^{ij} \bar\eta_i  \Lambda_j=- \nabla_m(\varepsilon^{ij}\bar\epsilon_i  \gamma^m\Lambda_j)+(2w-4)\varepsilon^{ij}\bar\eta_i\Lambda_j\,,
}}
where as in the previous section $\bar\lambda=\lambda^T C$.
 The ${\cal N}=2$ superconformal invariant realizing an exactly marginal operator is  constructed from such a multiplet of Weyl weight $2$. Indeed, it follows from    \SUSY\ that 
 \eqn\killing{
I=\int d^4x \sqrt{g}\, C\, 
}
is invariant, where $C$ is the top component of the multiplet~\SUSY.

The complex parameters that span the conformal manifold and source the  ${\cal N}=2$  exactly marginal operators  are realized as the bottom components of $\CN=2$ chiral multiplets of Weyl weight $0$. Note that Weyl weight 0 chiral superfields are irreducible~\deRooMM, thus, in general, we cannot embed the exactly marginal couplings in $\CN=2$ vector multiplets.     
  
 We are interested in unraveling the physical content of the sphere partition function of $\CN=2$ SCFTs.  We consider regulating the ultraviolet divergences  in an $OSp(2|4)$ invariant fashion. This is the  ${\cal N}=2$ supersymmetry algebra of a massive theory on the four-sphere $S^4$.  We describe it below in detail. We will prove that the $S^4$ partition function of $\CN=2$ superconformal field theories regulated in an  $OSp(2|4)$ invariant fashion  computes the K\"ahler potential on the conformal manifold
 \eqn\kahlerforu{
 Z_{S^4}=e^{{K}/12}\,.}
  
   The $OSp(2|4)$ supersymmetry transformations are generated by  conformal Killing   spinors $\epsilon^j$ and $\epsilon_j$ obeying
\eqn\killing{
\nabla_m \epsilon^j={i\over 2r} \gamma_m \tau_1^{jk} \epsilon_k\qquad \nabla_m \epsilon_j={i\over 2r} \gamma_m \tau_{1 jk} \epsilon^k\,,}
where $\tau_a^{jk}=(i \sigma_3,-1,-i \sigma_1)=({\tau_a}_{jk})^*$, and $\sigma_a$ are the   Pauli matrices. These correspond to  conformal Killing spinors  \ckstwo\ with 
\eqn\confs{
\eta^j={i\over 2r} \tau_1^{jk} \epsilon_k\qquad \eta_j={i\over 2r} \tau_{1 jk} \epsilon^k\,.}
We can diagonalize    equations \killing\ by defining
\eqn\diag{
\chi^j =\epsilon^j+   \tau_1^{jk}\epsilon_k\qquad \hat \chi^i=\epsilon^j-   \tau_1^{jk}\epsilon_k\ \,,}
so that 
\eqn\killingdiag{
\nabla_m \chi^j={i\over 2r} \gamma_m \chi^j\,.}
and 
  $\hat \chi^j=\gamma_* \chi^j$, where $P_{L/R}={1\over 2}(1\pm \gamma_*)$. 
  
  Computing the commutator of the ${\cal N}=2$ superconformal transformations   \SUSY\ acting on the fields of a  chiral multiplet gives a representation of the  ${\cal N}=2$ superconformal algebra on the fields
\eqn\confsugra{
[\delta_1,\delta_2]= \xi^m P_m+\lambda_a R^a+\lambda _{{\rm D}} D+\lambda _{R} R+ {1\over 2}\lambda^{ab} L_{ab}\,.}
With the choice of spinors \killing, it is easy to prove that the vector field
produced by two superconformal transformations
\eqn\killingc{
\xi_m={1\over 2} \bar\epsilon_2^i \gamma_m \epsilon_{1i}+{1\over 2}\bar\epsilon_{2 i} \gamma_m \epsilon_1^i}
is a Killing vector on $S^4$. Moreover, with the choice of spinors \killing\ the  parameters associated to local dilatations and    $U(1)_R$ R-symmetry vanish, while    the $SU(2)_R$ R-symmetry is broken down to $SO(2)_R$
\eqn\breakconfdos{\eqalign{
\lambda _{{\rm D}}=& - {1\over 2}\left( \bar \epsilon ^i_1\eta _{2i}+\bar \epsilon _{1i}\eta _2^i-    \bar \epsilon ^i_2 \eta _{1i}
-\bar \epsilon_{2i} \eta ^i_1\right)=0~,\cr
\lambda _{R} =&  {i\over 2}\left( \bar \epsilon ^i_1\eta _{2i}-\bar \epsilon _{1i}\eta _2^i -\bar \epsilon ^i_2\eta _{1i}+\bar \epsilon _{2i}\eta _1^i\right)=0~,\cr
 \lambda_a=& \lambda _{j}{}^i\tau_{ai}{}^j= (-\bar \epsilon ^i_1\eta _{2j}+\bar \epsilon _{1j}\eta _2^i+\bar \epsilon ^i_2\eta _{1j}-\bar \epsilon _{2j}\eta _1^i)\tau_{ai}{}^j\Longrightarrow  \lambda_1  ={i\over r}\varepsilon_{ij}\bar\chi_{1}^i\chi_{2}^j\,,  ~~ \lambda_2=\lambda_3=0\,.}}
Thus the  Killing spinors generate the $OSp(2|4)$ supersymmetry algebra, whose supercharges close into an $SO(5)$ isometry and an $SO(2)_R$ R-symmetry.\foot{On the fields, a   local Lorentz transformation with  parameter $\lambda^{ab}=-\nabla^{[a}\xi^{b]}$ is also induced.}
  This is the symmetry of a general   massive ${\cal N}=2$   theory on $S^4$.

Using the $OSp(2|4)$ supersymmetry transformations of a chiral multiplet  \SUSY\ we expect that the top component $C$ of the multiplet with Weyl weight $2$ can be written as three consecutive supersymmetry transformations of a linear combination of fermions in the multiplet modulo a zero and a total derivative. One could then repeat the argument of section~3 and arrive at~\kahlerforu. Instead, here we follow a closely related strategy, extending  to four dimensions the localization proof of~\kahlerforu\ presented in \GomisWY.  In addition, we derive~\kahlerforu\ by an explicit supersymmetric regularization. The two derivations agree. 
  
   \subsec{The K\"ahler Potential from the Four-Sphere}

 First, we  employ  the localization computation \PestunRZ\  of the $S^4$  partition function of Lagrangian four dimensional ${\cal N}=2$ superconformal field theories. These theories are based on vector multiplets with gauge group $G=\prod_i G_i$ and hypermultiplets transforming in a representation $R$ of $G$. The partition function can be  computed by localizing the functional integral with respect to a supercharge in $OSp(2|4)$. For our analysis, the details of the hypermultiplets, which vanish on the localization saddle points \PestunRZ, are not important. Therefore,  we can focus on the ${\cal N}=2$  vector multiplets. 
 
  The $\CN=2$ supersymmetric vector multiplet action is constructed from the top component of an ${\cal N}=2$  chiral and anti-chiral multiplet of Weyl weight 2 
   \eqn\sym{
 A_i={i\over 2}F(\Phi_i)= {i \over 8\pi}\hbox{Tr}\,\Phi_i^2\qquad 
\bar A_i=-{i\over 2}\bar F(\bar\Phi_i)=-{i \over 8\pi}\hbox{Tr}\,\bar \Phi_i^2~,
}
where $\Phi_i$ is an ${\cal N}=2$ vector multiplet of Weyl weight 1 associated to the gauge group factor $G_i$.\foot{The argument below can be carried without referring to a  specific microscopic realization. This requires solving for the supersymmetric configurations of the $\CN=2$ chiral multiplet of Weyl weight 2. To simplify the analysis we refer to the elementary fields $\Phi_i$.}
The four-sphere  $\CN=2$ supersymmetric vector multiplet action is
 \eqn\optionact{
S=\int d^4x\sqrt{g}\sum_i \left({\tau_i}\, C_i +{\bar\tau_{\bar i}}\, \bar C_{\bar i}\right)\,,}
where $C_i$ and $\bar C_{\bar i}$ are the top components of the  composite chiral and anti-chiral multiplets.
The exactly marginal parameters are the complexified gauge  couplings 
\eqn\coupl{
\tau_i={\theta_i\over 2\pi}+{4\pi i\over g_i^2}\,.}

 Calculating the second derivative with respect to the marginal couplings we get
 \eqn\deriva{
 \partial_i\partial_{\bar j} \log Z_{S^4}={1\over \pi^4}  \vev{\int_{S^4} d^4x\sqrt{g}\, C_i(x)\ \int_{S^4} d^4y\sqrt{g}\, \bar C_{\bar j}(y)}
 =(32 r^2)^2\vev{A_i(N) \bar A_{\bar j}(S)}.}
 In the final step, we have used supersymmetric localization.
 
 To relate this to the Zamolodchikov metric we can use the supersymmetry transformations~\SUSY\ to relate the two-point function of the bottom components~\deriva\ to the two-point function of the top components.  
 This yields $\vev{A_i(N) \bar A_{\bar j}(S)}={r^4\over 48}\vev{C_i(N) \bar C_{\bar j}(S)}$. Finally, we need to relate the two-point function $\vev{C_i(N) \bar C_{\bar j}(S)}$ to the Zamolodchikov metric. The result is $\vev{C_i(N) \bar C_{\bar j}(S)}={1\over (2r)^8}g_{i\bar j}$. Combining all the factors we find
 \eqn\derivai{
 \partial_i\partial_{\bar j} \log Z_{S^4}= {1\over 12}g_{i\bar j}={1\over 12}\partial_i\partial_{\bar j} K~. }
 This shows  \maini
 \eqn\finalKK{
 Z_{S^4}=e^{K/12}\,.}
 
We can now compare this result to another derivation. In section~3, equation~\integra, we have evaluated the integrated two-point function using regularization by analytic continuation in the scaling dimension. As we have explained in section~3, this regularization does not always work. One needs to make sure that it preserves the massive supersymmetry algebra. In the context of $\CN=2$ on $S^4$, this is indeed the case, because one can vary the scaling dimension of a chiral operator without breaking the $SO(2)_R\subset SU(2)_R$ symmetry. Plugging $d=4$ in~\integra\ and taking the limit $\epsilon\rightarrow 0$, we find precisely~\derivai\ (after correctly normalizing the operators, as in~\deform). This therefore provides a derivation of~\derivai\ that does not depend on localization. 

Let us discuss a simple example, that of $\CN=4$ super-Yang-Mills on the four-sphere. The conformal manifold has one complex parameter that preserves $\CN=4$ supersymmetry, i.e. the gauge coupling. The $S^4$ partition function depends on the masses of the fields in the adjoint hypermultiplet, and those need to be tuned such that the four-sphere partition function is that of the conformally coupled theory. For the correct, conformal  choice of the mass parameter  \OkudaKE, the instanton contributions to the four-sphere partition function vanish, and one finds via our prescription~\finalKK\ $K\sim \log(i(\bar\tau-\tau))$. If we perform an $SL(2,Z)$ $S$-duality transformation, the K\"ahler potential shifts by $\log(\tau)+\log(\bar \tau)+const$. This is a manifestation of a fact that we have explained in detail in section~3; the partition function is a section rather than a function. This K\"ahler transformation should be understood as a local supergravity counterterm, similar to the ones we have found in section~3. We leave this as well as the  study of the K\"ahler geometry in more general $\CN=2$ examples for the future.

\centerline{\bf Acknowledgments}
We are very grateful to  C.~Closset,  L.~Di~Pietro, N.~Doroud, T.~Dumitrescu, B. Le Floch, D.~Gaiotto, S.~Lee, J.~Maldacena, V.~Narovlansky, J.~Polchinski, A.~Schwimmer, A. Van Proeyen 
for useful discussions. We especially thank  N.~Seiberg for helpful discussions at various stages of the project.  JG and ZK are  grateful to the KITP for its warm hospitality during the initial stages of this project, which was supported in part by the National Science Foundation under Grant No. NSF PHY11-25915. EG and ZK thank the Perimeter Institute for its very kind hospitality during the course of this project. Research at the Perimeter Institute is supported in part by the Government of Canada through NSERC and by the Province of Ontario through MRI. J.G. also acknowledges further support from an NSERC Discovery Grant and from an
ERA grant by the Province of Ontario. ZK  is supported by the ERC STG grant number 335182, 
by the Israel Science Foundation under grant number 884/11, 
by the United States-Israel Binational Science Foundation (BSF) under 
grant number 2010/629, 
and by the I-CORE Program of the 
Planning and Budgeting Committee and by the Israel Science Foundation under 
grant number 1937/12.  
Any opinions, findings, and conclusions or recommendations expressed in this material 
are those of the authors and do not necessarily reflect the views of the funding agencies. 

\listrefs

\end